# Design principles for amorphous solid-state electrolytes

Qifan Yang[1, 2], Xiao Fu[1, 2], Xuhe Gong[1, 4], Jingchen Lian[1, 3], Liqi Wang[1, 3], Ruijuan Xiao[1, 2, 3, *], Yong-Sheng Hu[1, 2, 3, *] and Hong Li[1, 2, 3, *]

[1] Beijing National Laboratory for Condensed Matter Physics, Institute of Physics, Chinese Academy of Sciences, Beijing 100190, China

[2] Center of Materials Science and Optoelectronics Engineering, University of Chinese Academy of Sciences, Beijing 100049, China

[3] School of Physical Sciences, University of Chinese Academy of Sciences, Beijing 100049, China

[4] School of Materials Science and Engineering, Key Laboratory of Aerospace Materials and Performance (Ministry of Education), Beihang University, Beijing 100191, China

E-mail: rjxiao@iphy.ac.cn, yshu@iphy.ac.cn, hli@iphy.ac.cn

**Abstract**

Amorphous solid-state electrolytes (SSEs) offer unique advantages for next-generation batteries, but their rational design is hindered by an unclear structure-property relationship. This study establishes universal design principles through atomistic simulations of 32 amorphous Li-M-X systems (M = B, Al, Si, P; X = F, Cl, Br, I, O, S, Se, N). We identify four structure types governed by a rule that saturated M-X groups with more negative charges preferentially form M-X-M chains, determine paddle-wheel and cooperative migration as two favorable transport mechanisms that are significantly enhanced in amorphous structures. We also pinpoint oxides/fluorides as optimal for electrochemical/hydrolytic stability, and reveal bulk modulus as a simple predictor for $Li^+$ mobility. These insights are integrated into a practical design diagram, providing a novel and valuable framework for advancing high-performance amorphous SSEs.

## Main

High-performance solid-state electrolytes (SSEs) have constantly been a key area of research for all-solid-state batteries[1]. Compared to extensively explored crystalline SSEs, such as NASICON-type $LiM_2(PO4)_3$ (M = Ge, Ti, Sn, Hf, Zr)[2], Li argyrodites[3], [4], garnet-type $Li_xLa_3M_2O_{12}$ ($5 \leq x \leq 7$, M = Nb, Ta, Sb, Zr, Sn)[5], [6], $Li_{10}GeP_2S_{12}$ (LGPS)[7], [8] and its high-entropy derivative $Li_{9.54}[Si_{0.6}Ge_{0.4}]_{1.74}P_{1.44}S_{11.1}Br_{0.3}O_{0.6}$[9], as well as $Li_{2-x}La_{(1+x)/3}M_2O_6F$ (M = Nb, Ta)[10], amorphous SSEs exhibit distinct advantages. These include isotropic high ionic conductivity, ease of synthesis in various shapes, and broader compositional variations possibility, etc.[11], making them highly promising for the next-generation solid-state batteries. However, due to insufficient understanding of the complex relationships between chemical composition, atomic-scale disorder, and resulting properties such as $Li^+$ ion transport property and mechanical property, a universal guiding principle for designing amorphous SSEs has yet to be established[12].

Unlike periodically ordered crystalline materials, amorphous systems universally exhibit short-range order and long-range disorder[13] - manifesting as metastable configurations with unique characteristics, resulting in special physical and chemical properties. This inherent atomic-scale disorder for amorphous materials establishes two critical consequences: (1) Computational challenges and solution: For long-range disordered amorphous systems, ab-initio molecular dynamics (AIMD) simulations, which is a relatively accurate method commonly used for crystalline systems, faces limitations due to finite-size effects and restricted simulation timescales. These constraints allow only qualitative assessment of ion migration mechanisms, while making accurate ionic conductivity calculations unfeasible[14]. Therefore, we utilize AIMD for preliminary structural analysis and ion migration mechanism exploration, and perform ionic conductivity computations via machine learning interatomic potential based molecular dynamics (MLIP-based MD) to extend both the simulation space and time scale. (2) A new structure-property paradigm: Different from crystals where structural variations can yield vastly diverse properties even at fixed compositions, amorphous materials exhibit a fundamental structural commonality due to their inherent atomic-scale disorder. It redefines the governing principles of property regulation. Rather than long-range order, short-range structural features become the key determinants of amorphous material behavior, including ion transport and mechanical properties. Consequently, in amorphous materials, elemental pairings and compositions become the primary levers for property control, as they directly dictate these local structural units and characteristics. It should be noted that the synthesis protocol for amorphous materials also plays a role in determining their structures and properties[15]. However, in simulations, by maintaining a consistent modeling process, researchers can focus on the structural property determining factors of elemental pairings and compositions for amorphous systems.

Clues regarding the effects of elemental pairings and compositions on the ion transport properties of amorphous materials can be found in several recent studies. For instance, several recently synthesized amorphous oxychloride or chloride, $x$$Li_2O$-$MCl_y$ (M = Ta or Hf, $0.8 \leq x \leq 2$, $y$ = 5 or 4)[16], $Li_{1.75}ZrCl_{4.75}O_{0.5}$[17], $MAlO_xCl_{4-2x}$ (MAOC, M = Li, Na, $0.5 < x < 1$)[18], $LiTaCl_6$[19], and $Li_{1.3}ZrN_{0.4}Cl_{4.1}$[20] have high $Li^+$ ionic conductivity on the order of mS/cm. Elemental combinations and compositional ratios affect ion migration in amorphous systems mainly through their decisive role in shaping local structural environments and the kinetic properties of these units. The computational modeling and simulations on amorphous $MAlO_xCl_{4-2x}$ structures indicated that Al atoms are connected into chains through O or Cl atoms (O atoms in most cases), forming the Al chain skeleton and the polymer-like structure, and their great ionic transport performance is attributed to the rotation of Cl atoms on Al-O-Cl polyanionic groups[21]. The theoretical research for amorphous $LiTaCl_6$ also verified the importance of the motion of Cl anions on $Li^+$ ions' transport[22]. Similarly, Reference [23] suggested that the widely researched amorphous Li-P-S system exhibits remarkable $Li^+$ ion transport ability related to the paddle-wheel mechanism - an ion conduction mechanism in which mobile ions are “pushed” or “facilitated” by the rotational motion of surrounding polyanionic groups. However, some studies suggest that the paddle-wheel mechanism in amorphous/crystalline Li-P-S systems does not exist[24] due to the mismatch between the large rotation frequency of polyanionic groups and the lithium ion hopping frequency. On the contrary, polyanionic groups adjust their orientation through the soft-cradle effect, indirectly promoting the transport of $Li^+$ ions[25]. Actually, the aforementioned dispute is related to the conceptual definition of the paddle-wheel mechanism, more specifically, the definition of rotation[26]. A recent study also revealed that vibrational modes of S atoms within $[PS_4]^{3-}$ tetrahedra - rather than rotational or translational motions - dominate the $Li^+$ conduction process in crystalline $Li_3PS_4$[27]. The existing studies suggest that specific elemental pairings (Al-Cl, P-S, and so on) and compositions in amorphous materials enable high ionic conductivity, with the anion dynamics critically influencing $Li^+$ ion transport. Thus, it is crucial to solve the unknown mysteries in amorphous SSEs, including: What are the atomic-scale structural characteristics of amorphous SSEs and how can they be regulated through specific elemental combinations and compositional design? What are the ion transport mechanisms in amorphous SSEs, and how do they correlate with structures and local bonding environments? What general trends govern their electrochemical stability, hydrolytic stability, and mechanical properties? Can ionic transport and related properties of amorphous SSEs be modulated and optimized through strategic tuning of elemental pairings and compositions? Addressing these questions through systematic research on the structural characteristics, ionic transport mechanisms, electrochemical and hydrolytic stability, and mechanical properties for amorphous materials will help to establish universal design principles for high-performance amorphous SSEs.

To explore these fundamental questions, we first constructed the structural models for 32 distinct amorphous Li-M-X systems (M = B, Al, Si or P; X = F, Cl, Br, I, O, S, Se or N) using AIMD simulations and density functional theory (DFT) calculations. Subsequently, we performed additional AIMD simulations on these models. M cations (light, geochemically abundant elements with +3 to +5 oxidation states) and X anions (spanning -1 to -3 oxidation states) were strategically selected with the goal of improving the gravimetric energy density and reducing the cost of lithium batteries. Critically, we maintained isolated M-X polyanionic groups in initial structures to exclusively probe elemental pairing effects on ion transport. The AIMD results explained the structural characteristics of amorphous Li-M-X systems and revealed the prevalent ion transport mechanisms for amorphous SSEs. In order to better illustrate the ionic transport mechanisms for amorphous materials, we selected example structures, amorphous $LiBF_4$ with paddle-wheel mechanism and $Li_4SiSe_4$ with $Li^+$ ions' cooperative migration mechanism, to analyse their ion transport behavior and calculate their ionic conductivity at 300 K using MLIP-based MD simulations. Also, the velocity autocorrelation power spectrum of $Li^+$ and $Cl^-$ in both crystalline and amorphous $LiAlCl_4$ was calculated to clarify the anion-cation coupling in amorphous systems. Then, by analyzing the $Li^+$ ion transport in both crystalline and amorphous $LiAlCl_4$ systems with different volumes, we determined the distinct critical volumetric conditions triggering the paddle-wheel mechanism in each type of structure. Beyond ionic mobility, we evaluated the electrochemical window, hydrolytic stability and bulk moduli for amorphous Li-M-X systems, revealing that these properties are primarily element-dependent, and investigated correlations between bulk modulus and $Li^+$ mobility for amorphous materials. Ultimately, a universal design diagram for atomically disordered SSEs is established according to the M-X elemental pairing and the variations in anion's type and concentration.

## Results

### The generation and simulation process for amorphous Li-M-X structures

To generate amorphous structures, the common method is high-temperature melting and quenching. First of all, as displayed in Figure 1a, we directly constructed the initial disordered Li-M-X structures, $Li_{cx-m}M^{m+}[X^{x-}]_c$ ($m$ = 3, 4, 5 and $x$ = 1, 2, 3 for M element and X element respectively, $n$ is the saturated coordination number for M-X polyanionic groups in crystals, and the atomic ratio $c$ is defined as $n$ if $nx > m$ or as the largest integer not exceeding $m/x$+1 if $nx \leq m$ to ensure charge neutrality), by randomly placing Li atoms and $MX_n$ polyanionic groups rather than from crystals, ensuring a consistent treatment across all 32 diverse systems. Crucially, the subsequent equilibration process was sufficient to drive each system into a well-defined local energy minimum, effectively erasing the memory of the specific crystalline starting point. This ensures that the structural models used for analysis are representative of the amorphous state for each composition, independent of the preparation pathway within the scope of this comparative study. For systems with significant oxidation-state disparities between M and X elements (e.g. amorphous Li-P-Cl system with $P^{5+}$ and $Cl^-$), isolated $X^{x-}$ anions need to be added to achieve valence balance, a scenario which occurs when the X/M atomic ratio $c$ exceeds the saturation coordination number $n$ of the M cation in M-X groups. Each simulation box contains ~ 80 atoms consistent with the approach used in Reference [28]. To isolate the impact of M-X pairings, the initial structures only contain unlinked M-X groups, minimizing interference from extended M-X-M chains in $Li^+$ ion conductivity. Meanwhile the diverse melting points of Li-M-X structures complicates the selection of melting temperatures. Excessively high simulated temperatures will lead to unreasonable structures, while excessively low temperatures may result in insufficient equilibrium. Consequently, we performed a two-step equilibration process using AIMD simulations: Process 1 (equilibration at 300 K for 40 ps) and Process 2 (melting at 1500 K followed by quenching to 300 K), as shown in Supplementary Figure S1. We acknowledge that the 100 K/ps quench rate used in Process 2 exceeds experimental conditions, a common trade-off in AIMD simulations for computational feasibility. However, the short-range order critical to ion transport is well preserved, and since our analysis relies on relative comparisons across the 32 systems rather than absolute quantitative values, this computational expediency is unlikely to introduce systematic bias that would affect our primary conclusions. We systematically compared key structural and thermodynamic parameters, including free energies, RDF profiles, and coordination number curves between these two processes (Supplementary Table S1 and Supplementary Figure S2). Results show that except nitrides, all other structures exhibit negligible free energy differences ($-0.06 \leq \Delta E \leq 0.06$ eV/atom, and for some structures the free energy is lower after Process 1, while for others it's lower after Process 2, without exhibiting a specific pattern), similar RDF curves and identical coordination number curves. The four oxides with higher melting points have already melted and reached energy equilibrium at 1500 K (Supplementary Figure S3), suggesting the successful amorphization for most materials except nitrides. However, for nitrides, the structures generated through Process 2 are systematically more stable by over 0.1 eV/atom than those from Process 1. These results suggest that direct equilibration at 300 K is viable for amorphization in most non-nitride systems studied in this work, while for refractory nitrides, elevated temperature processing is required to achieve glassy states. Accordingly, as Figure 1b illustrates, the nitride systems were melted at higher temperatures (1900 K for $Li_3BN_2$ and $Li_9AlN_4$; 2500 K for $Li_8SiN_4$ and $Li_7PN_4$). The MSDs and energy evidence from AIMD simulations shown in Supplementary Figure S4 confirm complete melting at corresponding temperatures prior to quenching. This comprehensive treatment ensures full structural equilibration of all systems. During the above structural equilibration steps, some polyanionic groups experienced bond-breaking in certain structures, leading to the formation of isolated X anions. In order to determine the equilibrium density of each amorphous $Li_{cx-m}M^{m+}[X^{x-}]_c$, we adopted DFT calculations to perform structural relaxation for both box size and atomic positions. Unlike crystalline phases, whose thermodynamic stability can often be evaluated using energy above the convex hull, amorphous structures are metastable states with higher energy, typically frozen from a disordered configuration. Therefore, we adopted MD simulations to assess the kinetic stability of each composition, and found that all final configurations were dynamically stable with no signs of crystallization. The methodology for building and equilibrating these amorphous Li-M-X structures is detailed in Supplementary Note S1. These amorphous structures were characterized systematically through the calculations of their electrochemical windows, hydrolytic stability, and bulk moduli to facilitate comprehensive property analysis. Furthermore, to study the effect of M-X pairings on the ionic transport properties, as suggested in Figure 1c, we conducted an additional 60 ps AIMD simulation at 300 K for amorphous Li-M-X structures. Finally, in the following sections, to evaluate the intrinsic ionic transport performance and calculate the ionic conductivity at 300 K of selected materials, we trained MLIPs by Neural Equivariant Interatomic Potentials (NequIP)[29] code and conducted MLIP-based MD simulations for 3000 ps (~500 atoms, 300 K). The training process for MLIP models and the MLIP-based MD simulation process are described in Supplementary Note S2, and the corresponding model performance is displayed in Supplementary Table S2. Direct simulations at 300 K were employed to avoid potential interpolation errors arising from high-temperature structural transformations (e.g., melting or glass transition) in amorphous systems.

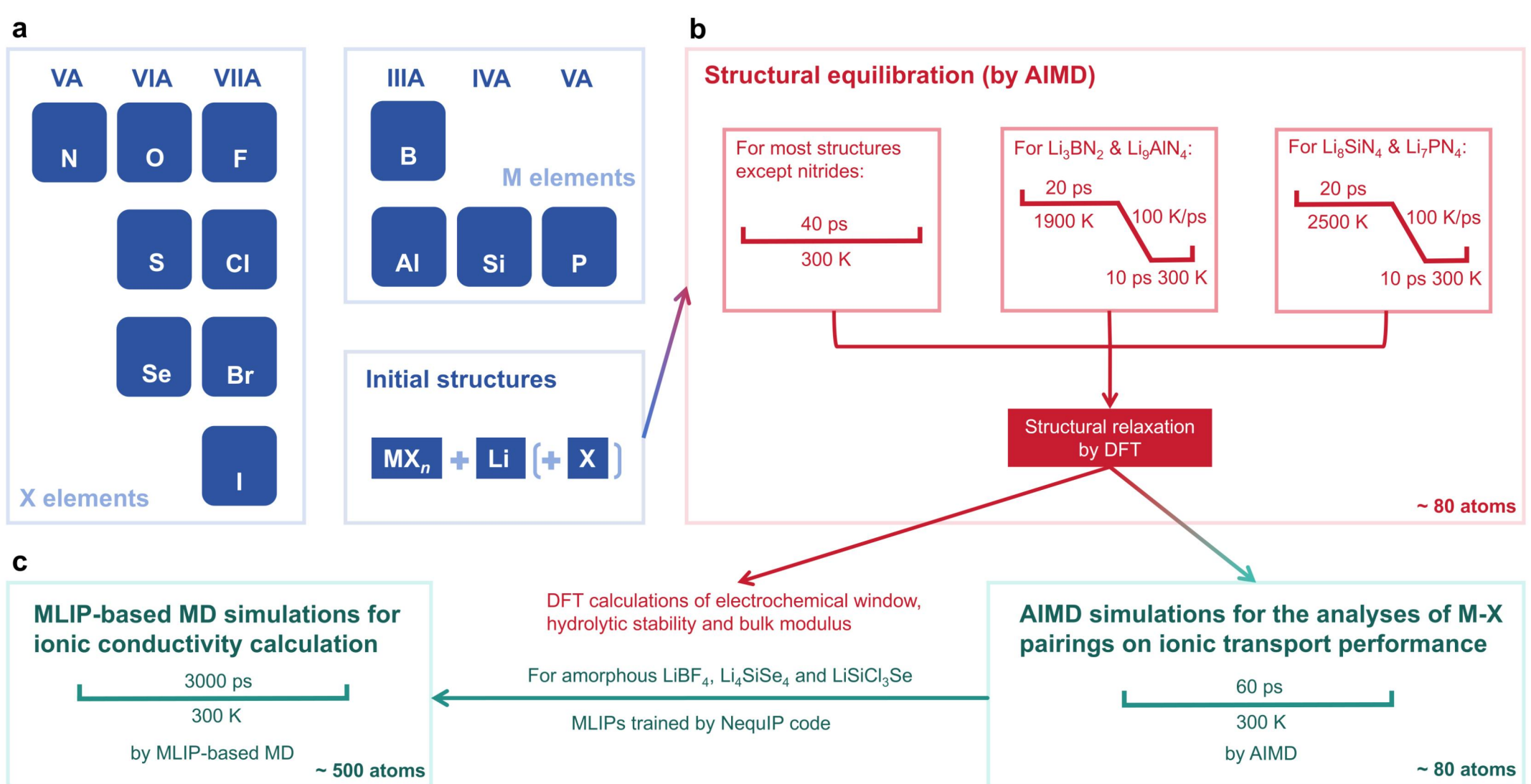

**Figure 1. The generation and simulation process for amorphous Li-M-X structures.**
**a** Element types of M cations and X anions in amorphous Li-M-X structures (M = B, Al, Si or P; X = F, Cl, Br, I, O, S, Se or N), and the settings of the initial disordered Li-M-X structures. **b** The structural equilibration steps for the initial disordered Li-M-X structures using AIMD simulations and DFT structural relaxation. **c** The AIMD simulations for amorphous Li-M-X structures and the MLIP-based MD simulations for selected amorphous structures.

**Structural characteristics of amorphous Li-M-X structures**

Based on atomic structure analysis of the constructed Li-M-X amorphous systems, we found that the amorphous structural frameworks can be classified into four types, despite the presence of mobile $Li^+$ ions. Type 1: structures containing both $MX_a$ polyanionic groups ($a \leq n$; e.g. $[PBr_3]^{2+}$ or $[PBr_4]^+$ with $n = 4$) and isolated X anions, where the latter either compensate for valence imbalance or arise from unsaturated local coordination formed by polyanionic group dissociation; Type 2: structures containing only saturated polyanionic groups, e.g. $[BF_4]^-$, without the presence of isolated X anions; Type 3: structures in which less than or equal to 50% of M cations associated with M-X-M chains, e.g. $[P_2S_7]^{4-}$; and Type 4: structures with more than 50% of M cations located in M-X-M chains, e.g. $[Al_2O_7]^{8-}$ or $[Al_4O_{13}]^{14-}$. From Type 1 to Type 4, the tendency toward forming extended M-X-M chains in the structure becomes increasingly evident. We classified 32 amorphous Li-M-X systems according to the atomic-scale features and summarized them in Figure 2, with representative structures as follows: amorphous $LiPBr_6$ as Type 1, amorphous $LiBF_4$ as Type 2, amorphous $Li_3PS_4$ as Type 3, and amorphous $Li_5AlO_4$ as Type 4. We observed that the structural type of the amorphous configurations is closely linked to the net charge number carried by the saturated $MX_n$ polyanionic groups (Figure 2). Firstly, when the net charge $\geq 0$, it necessitates the introduction of isolated X anions to fulfill valence equilibrium, producing Type 1 structures. Moreover, essentially all structures with the net charge $\geq 0$ contain unsaturated $MX_a$ ($a < n$) polyanionic groups. Secondly, structures with net charges ranging from -1 to -4 present in all four types. This is because the size limitation of the structural models means there are no more local structures of polyanionic groups available for statistical analysis, resulting in these structures not being clearly distinguished by four different types. However, as shown in Figure 2, the overall trend of "the more negative the net charge carried by $MX_n$, the easier it is for the structure to form M-X-M chains" is evident. Finally, polyanionic groups with net charge $\leq -5$ tend to form extended M-X-M chain structures (Type 4 structures). This means that increasingly negative valences of saturated $MX_n$ polyanionic groups (adding X anions with a more negative valence) drive extended M-X-M chain formation, which explains why in Li-Al-O-Cl amorphous materials, Al atoms tend to be connected via O atoms more easily than through Cl atoms[18]. The net charge of saturated $MX_n$ affects the number and distribution of $Li^+$ ions in structures, leading to this pattern. Specifically, when net charge of $MX_n \geq 0$, electrostatic repulsion between $MX_n$ groups and positively charged $Li^+$ ions in the surrounding environment further destabilizes the local structure, leading to partial dissociation of $MX_n$ groups where some M-X bonds break and release X anions. This configuration substantially impedes M-X-M bond formation. As the net charge of $MX_n$ becomes negative, it means that the lithium ions gradually become evenly distributed and the local concentration increases, which weakens the repulsive force among negative polyanionic groups and spontaneously connects them together. The above findings are crucial for modulating amorphous SSEs, as the formation of M-X-M chains is likely to contribute to the improvement of glass forming ability[21].

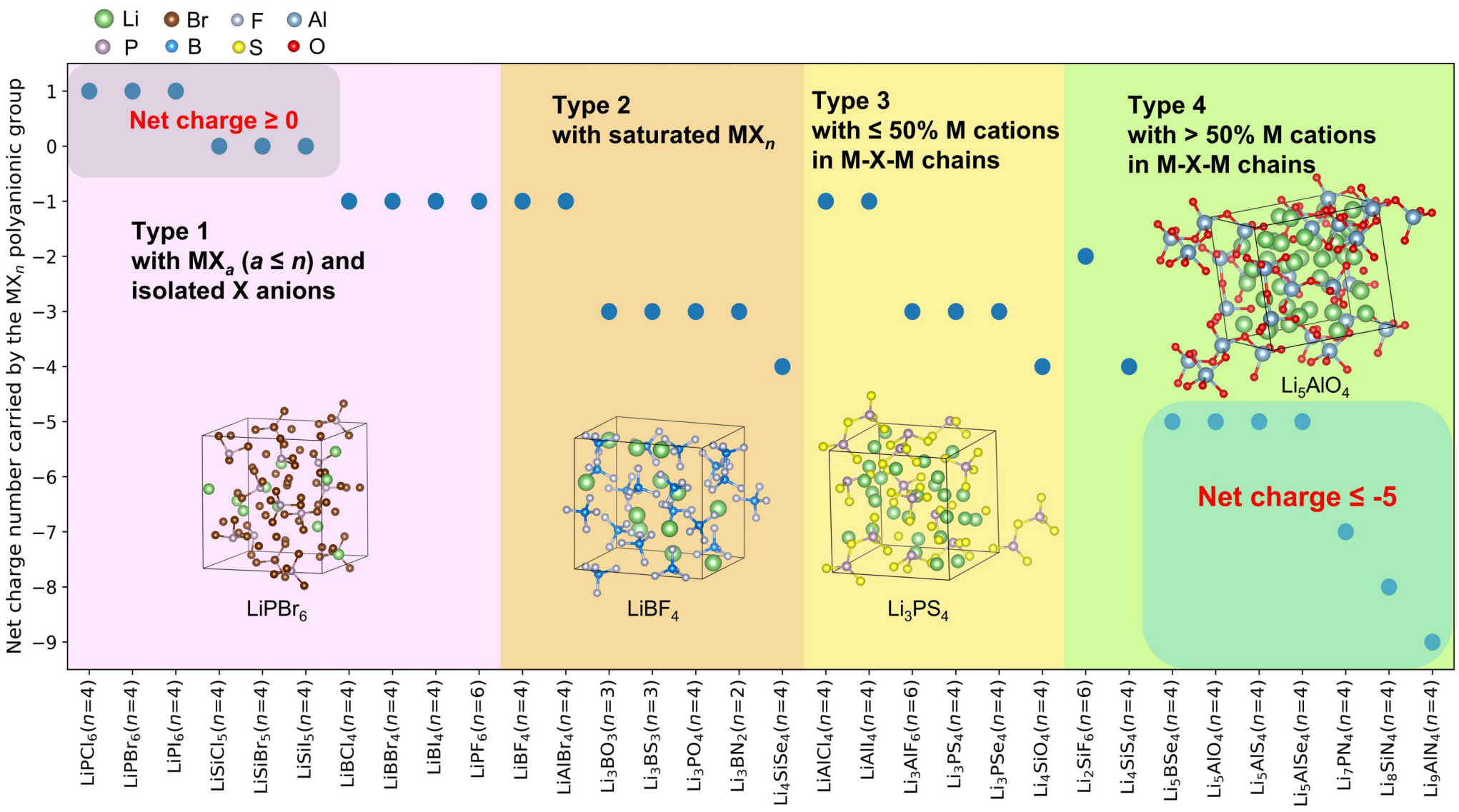

**Figure 2. The 4 types of amorphous Li-M-X structures and their relationship with net charge number carried by the saturated MX$_n$ polyanionic groups.**

**$Li^+$ ion transport properties for amorphous Li-M-X structures**

In order to investigate the $Li^+$ ion transport properties of amorphous structures with varying M-X pairings, the mean squared displacements (MSDs) of each element were calculated and compared based on the last 50 ps trajectory extracted from 60 ps AIMD simulations performed at 300 K. Figure 3a shows the MSD of Li element in 32 amorphous structures. The results demonstrate that the migration dynamics of $Li^+$ ions are highly pronounced in most Li-M-$X^-$ structures, exhibiting MSD > 1 Å$^2$ (except $Li_3AlF_6$), with a number of $Li^+$ ions in these structures experiencing long-distance migration during AIMD simulations, particularly significant when the $X^-$ ion is $Cl^-$, $Br^-$ or $I^-$. And for Li-M-$X^{2-}$, there are eight systems of all sulfides and selenides exhibiting Li MSD values exceeding 1 Å$^2$. Moreover, $Li^+$ diffusion is not evident in any Li-M-$X^{3-}$ systems. Figure 3b shows the MSD of element M. For amorphous materials with $X^-$ ions being $Cl^-$, $Br^-$, or $I^-$, the M cations exhibit significant MSD, indicating that the framework exhibits strong deformation capability under these compositions. Figure 3c and 3d display the MSD of X element and the percentage of X anions with MSD > 1 Å$^2$ respectively, which represent the X anions' mobility. In these systems, when the MSD of X anions exceeds 1 Å$^2$, significant rotational motion of the X anions is confirmed, as demonstrated by the pronounced $F^-$ anion rotation in amorphous $LiBF_4$ (with $c = n$) and $Br^-$ anion rotation in amorphous $LiPBr_6$ (with $c > n$) (Supplementary Figure S5). The analysis of the correlation between $Li^+$ ion and X anion mobility in these 32 amorphous structures are presented in Figure 3e and 3f. Firstly, for Li-M-$X^-$ systems, most of them display excellent $X^-$ anions' mobility, and structures with stronger $X^-$ anions mobility demonstrate better lithium ion migration, while $Li_3AlF_6$ system with relatively immobile anions show negligible lithium ion migration. This phenomenon can be attributed to the paddle-wheel mechanism of M-X polyanionic groups (Figure 3g), where stronger rotational dynamics facilitate efficient $Li^+$ ion conduction. Among Li-M-$X^{2-}$ structures, sulfides and selenides have favorable lithium-ion transport capabilities while exhibiting minimal or negligible $X^{2-}$ anion movement (< 1 Å$^2$). We conducted an analysis of the individual lithium ion motion in these structures. The existing of similar MSD curves for neighbored $Li^+$ ions in each structure shown in Supplementary Figure S6 indicates that cooperative migration mechanism for lithium ions dominates in Li-M-$X^{2-}$ systems (Figure 3g). Additionally, although X anion mobility in these structures is relatively limited, inspired by previous studies[27], we investigated the potential correlation between Li-ion migration and anion vibrational dynamics. As depicted in Supplementary Figure S7, in our investigation of amorphous $Li_3PSe_4$, we conducted five parallel AIMD simulations with all Se anions fixed in position. Comparative analysis of the Li MSD between the fixed and unfixed structures revealed a contrast: in AIMD simulations, the fixed configurations exhibited complete absence of Li ion motion, while Li mobility was observed in the unfixed structure, indicating that the vibration of $X^{2-}$ anions is the reason for promoting the $Li^+$ ions' cooperative migration. This indicates that the $Li^+$ ionic conductivity of amorphous Li-M-$X^{2-}$ structures is sensitive to the concentration of lithium ion, as the cooperative migration mechanism becomes feasible only when the lithium content reaches a suitable and sufficiently high level to enable close proximity between adjacent $Li^+$ ions. In Reference [30], the ionic conductivity of the amorphous Li-P-S system increases by two orders of magnitude as the $Li_2S$:$P_2S_5$ ratio rises from 60:40 to 70:30, highlighting the strong dependence on lithium ion concentration as we predicted. In contrast, in systems where $Li^+$ transport is primarily facilitated by the high mobility of $X^-$ anions, the reliance on $Li^+$ ions concentration is minimal. This is evidenced by the similar $Li^+$ ionic conductivity (of the same order of magnitude) even as the $Li_2O$:$TaCl_5$ ratio rises from 52:48 to 64:36[16]. Additionally, no Li-M-$N^{3-}$ systems exhibit evident $Li^+$ or $N^{3-}$ ions diffusion. Here we can summarize that the general ion transport behaviors in amorphous SSEs are basically triggered by the interaction of the dynamics of X anions and $Li^+$ ions (Figure 3g): for Li-M-$X^-$ systems, such as amorphous $LiAlCl_4$, the vigorous mobility of $X^-$ anions produces the high $Li^+$ ionic conductivity by paddle-wheel mechanism; for Li-M-$X^{2-}$ systems (e. g. $Li_3PS_4$), some M-$X^{2-}$ with suitable lithium ion concentration can generate cooperative migration for neighboring $Li^+$ ions through the vibration of $X^{2-}$ anions to obtain high $Li^+$ ionic conductivity; Li-M-$N^{3-}$ structures lack enough lithium ion transport capability, as the $N^{3-}$ ions remain stable.

In addition to being part of polyanionic groups bonded to M cations, anions also can exist in an isolated state, particularly in Type 1 structures. Our prior study revealed that in crystalline frameworks, stable isolated anions induce structural frustration and disordered lithium-ion distribution. This creates spherical potential fields that facilitate cage transport around the anions, ultimately enabling high ionic conductivity via the isolated-anion mechanism (Figure 3g)[31]. Inspired by these findings, we explored the role of isolated anions in amorphous structures. Take amorphous $LiPBr_6$ as an example, where 61.67% $Br^-$ ions are coordinated to $P^{5+}$ in tetrahedral geometries, and the remaining 38.33% are isolated $Br^-$ anions. The former bonded to M cations exhibit rotation around M cations and remain no long-distance migration (as demonstrated in Supplementary Figure S8a and S8b, the P-Br bond length remains stable and no bond-breaking phenomenon is observed). However, isolated $X^-$ anions display long-distance migration, as evidenced by the transport path of an individual isolated $Br^-$ anion presented in Supplementary Figure S8c. Unlike the cage transport for $Li^+$ ions brought by isolated anions in crystals, no analogous cage for $Li^+$ ions around isolated $X^-$ anions was observed, because the highly dynamic motion of isolated $X^-$ anions cannot establish a stable spherical potential field around them. Besides, the long distance migration of isolated $Br^-$ anions can compromise the structural integrity of the amorphous framework, leading to cascading ionic motion, including that of lithium ions. Therefore, given that isolated $X^-$ anions introduce structural instability risk, their inclusion should be minimized in amorphous SSE design. For amorphous Li-M-$X^{2-}$ structures, isolated $X^{2-}$ anions remain nearly unmoved, and cage shaped transport around these anions appears in structures with lithium ion transport such as amorphous $Li_4SiS_4$ (Supplementary Figure S8d). This indicates that isolated $X^{2-}$ anions that remain stationary in amorphous materials can trigger the structural frustration as in crystals. However, due to the lack of ordered arrangement among these isolated anions in amorphous configurations, the cage transport pathways between them are not always interconnected, thereby limiting their overall contribution to ionic conductivity. Notably, studies on the correlation between the polyanion environment and ionic conductivity in amorphous Li-P-S superionic conductors also support the above analysis[28], in which the isolated $S^{2-}$ anions do not significantly enhance $Li^+$ ion conductivity in Li-P-S amorphous materials compared with other local environments. That is to say, structural frustration arises not only from the existence of isolated $X^{2-}$ anions but also need a suitable local environment, such as the symmetry of isolated anion's local structures, the arrangement and the types of isolated anions, etc.[31]. For amorphous Li-M-$X^{3-}$ structures, they exhibit negligible $Li^+$ and $N^{3-}$ mobility and are therefore excluded from this analysis. Despite the fact that the nitrides discussed in this work do not exhibit ionic conductivity, they can serve as stable structural frameworks for amorphous materials, particularly in promoting the formation of M-X-M chains, as demonstrated by recent studies on the amorphous Li-Zr-N-Cl system[20].

Therefore, we have revealed three major $Li^+$ ion transport mechanisms in amorphous structures: paddle-wheel mechanism for amorphous Li-M-$X^-$ systems, cooperative migration mechanism for amorphous Li-M-$X^{2-}$ systems and isolated-anion mechanism for systems containing non-bonding anions. Simultaneously, we have elucidated the formation conditions for the paddle-wheel mechanism - the existence of $X^-$ anions - in amorphous materials, offering new insights surrounding this phenomenon. To further illustrate the paddle-wheel mechanism and the cooperative migration mechanism in amorphous systems, we performed MLIP-based MD simulations for amorphous $LiBF_4$ and $Li_4SiSe_4$ at 300 K, as shown in Supplementary Figure S9. The simulation results of amorphous $LiBF_4$ at 300 K (Supplementary Figure S9b) are similar to those of amorphous Li-Al-O-Cl systems[21]. The motion of B represents the collective movement of $[BF_4]^-$, while the relatively large MSD of F and Li indicates their correlation, reflecting the paddle-wheel mechanism. The calculated room-temperature ionic conductivity of amorphous $LiBF_4$ is 4.33 mS/cm. However, in amorphous $Li_4SiSe_4$ with the cooperative migration mechanism induced by the vibration of $Se^{2-}$ anions, $Li^+$ ions move violently, yet both $Si^{4+}$ and $Se^{2-}$ remain relatively stable and without significant movement during the simulation (Supplementary Figure S9e). The calculated room-temperature ionic conductivity of amorphous $Li_4SiSe_4$ is 12.7 mS/cm. Additionally, the distinct $Li^+$ transport mechanisms in the amorphous materials are further elucidated by the van Hove correlation function $G_d\,(t,r)$, which gives the average density (the radial distribution) of other $Li^+$ ions at distance $r$ after a time $t$ with respect to the initial reference ion[32]. As shown in Supplementary Figure S9c, the $G_d\,(t,r)$ for amorphous $LiBF_4$ varies little and stays approximately at 1. The absence of significant spatical-temporal correlation indicates that $Li^+$ ion migration in this material is not cooperative. In contrast, the $G_d\,(t,r)$ for amorphous $Li_4SiSe_4$ in Supplementary Figure S9f demonstrates the pronounced signature of cooperative migration. Within a short distance of $r < 1$ Å, $G_d(t, r)$ rises to a value near to 4, indicating that the motion of $Li^+$ ions is highly correlated. The comparison clearly elucidates the differences between the two transport mechanisms presented in the amorphous materials.

Moreover, we also calculated the velocity autocorrelation power spectrum of $Li^+$ and $Cl^-$ in crystalline/amorphous $LiAlCl_4$, as Supplementary Figure S10 illustrates. The results showed that in the low-frequency region corresponding to vibration/rotation/translation, especially 0-1.5 THz, the spectrum density of $Li^+$ and $Cl^-$ in amorphous $LiAlCl_4$ was higher, and the two overlapped with each other. The anharmonic thermal librations of $Cl^-$ anions in the amorphous system have the potential to transfer momentum to $Li^+$ cations, thereby generating a driving force for lithium migration, proving the coupling of anions and cations, consistent with a paddle-wheel mechanism in amorphous $LiAlCl_4$. This result further proves the important role of anion-cation coupling in the migration of lithium ions in amorphous materials.

Except for the influence of elemental pairing on the $Li^+$ ionic transport of amorphous structures, we further investigated the local atomic structural conditions required to trigger the paddle-wheel mechanism since anion rotations necessitate sufficient free volume[26]. By comparing the $Li^+$ migration ability in crystalline and amorphous $LiAlCl_4$ systems, our analysis demonstrates that amorphous materials exhibit anion mobility at much smaller volumes compared to their crystalline structures, attributed to the inherent structural frustration present in amorphous systems introduced by atomic disorder (Supplementary Figure S11). This means that even at the comparable local spatial constrains, the paddle-wheel mechanism is easier to operate in amorphous systems compared with crystals. These results demonstrate that both volume expansion and atomic disorder enhance anion mobility, thereby promoting lithium-ion transport, making the anion-assisted $Li^+$ ion transport more prevalent in amorphous structures.

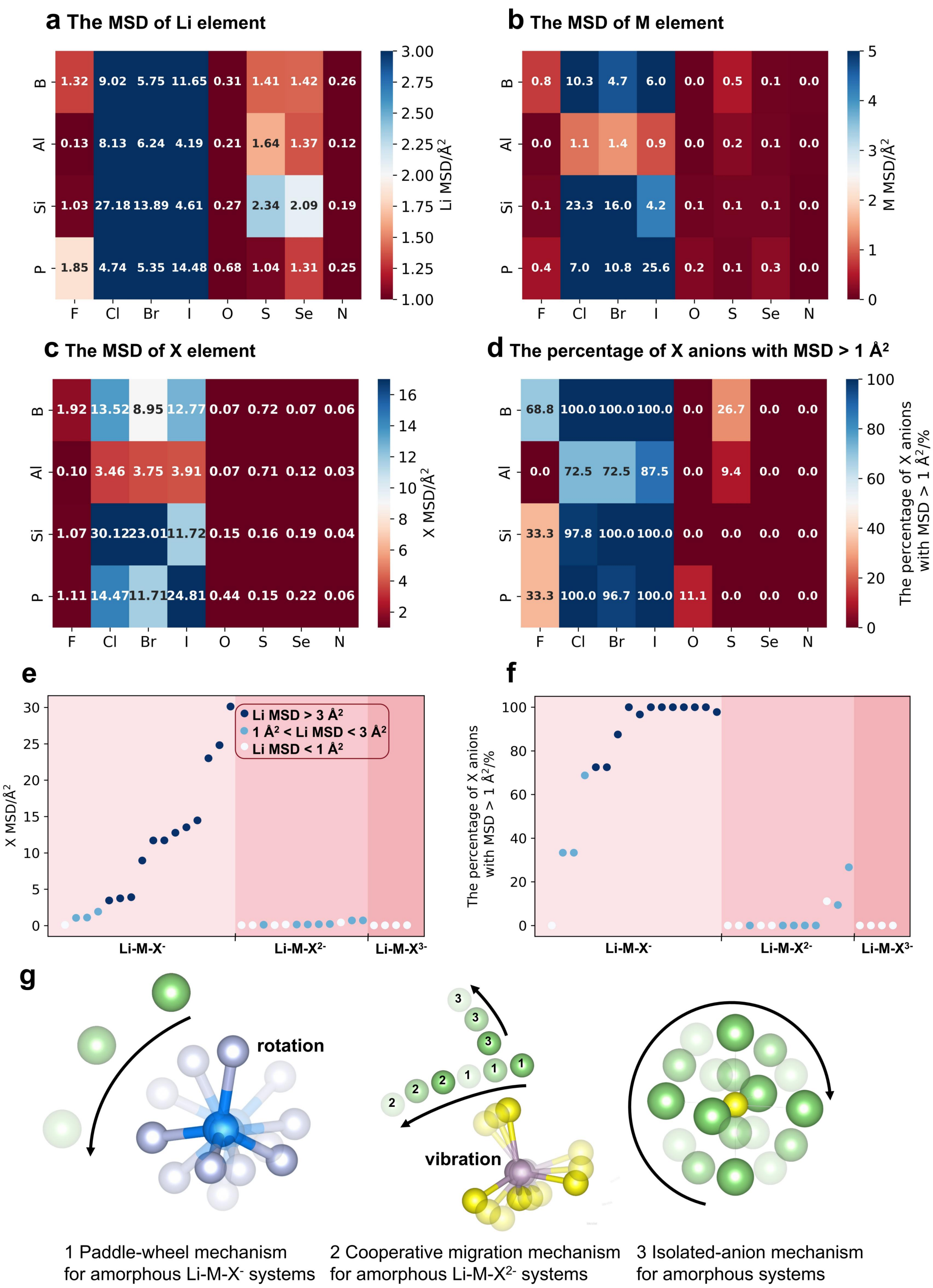

**Figure 3. The analysis of $Li^+$ ion transport characteristics for amorphous Li-M-X structures.** **a** MSD of Li element (Li MSD) calculated from 50 ps AIMD simulations at 300 K. **b** MSD of M element (M MSD) calculated from 50 ps AIMD simulations at 300 K. **c** MSD of X element (X MSD) calculated from 50 ps AIMD simulations at 300 K. **d** The percentage of X anions with MSD > 1 $Å^2$ calculated from 50 ps AIMD simulations at 300 K. **e** The correlation between X MSD and Li MSD for amorphous structures. **f** The correlation between the percentage of X anions

with MSD > 1 Å$^2$ and Li MSD for amorphous structures. **g** Three types of $Li^+$ ion transport mechanisms in amorphous materials.

### Electrochemical and hydrolytic stability for amorphous Li-M-X structures

Promising SSEs require not only high ionic conductivity, but also robust electrochemical stability and hydrolytic stability. To further evaluate the electrochemical and hydrolytic stability of amorphous Li-M-X structures, we calculated their electrochemical windows obtained by the giant potential phase diagram and reactivity with water obtained by the reaction diagram[33], shown in Figure 4. The results indicate that fluorides and oxides exhibit the widest electrochemical windows, followed by chlorides, consistent with trends observed in crystalline materials[34]. Additionally, some fluorides and oxides also demonstrate superior hydrolytic stability, while sulfides and selenides exhibit resistance to hydrolysis only when the cation M is $Si^{4+}$. Therefore, from a stability perspective, amorphous fluorides and oxides are optimal choices for SSEs.

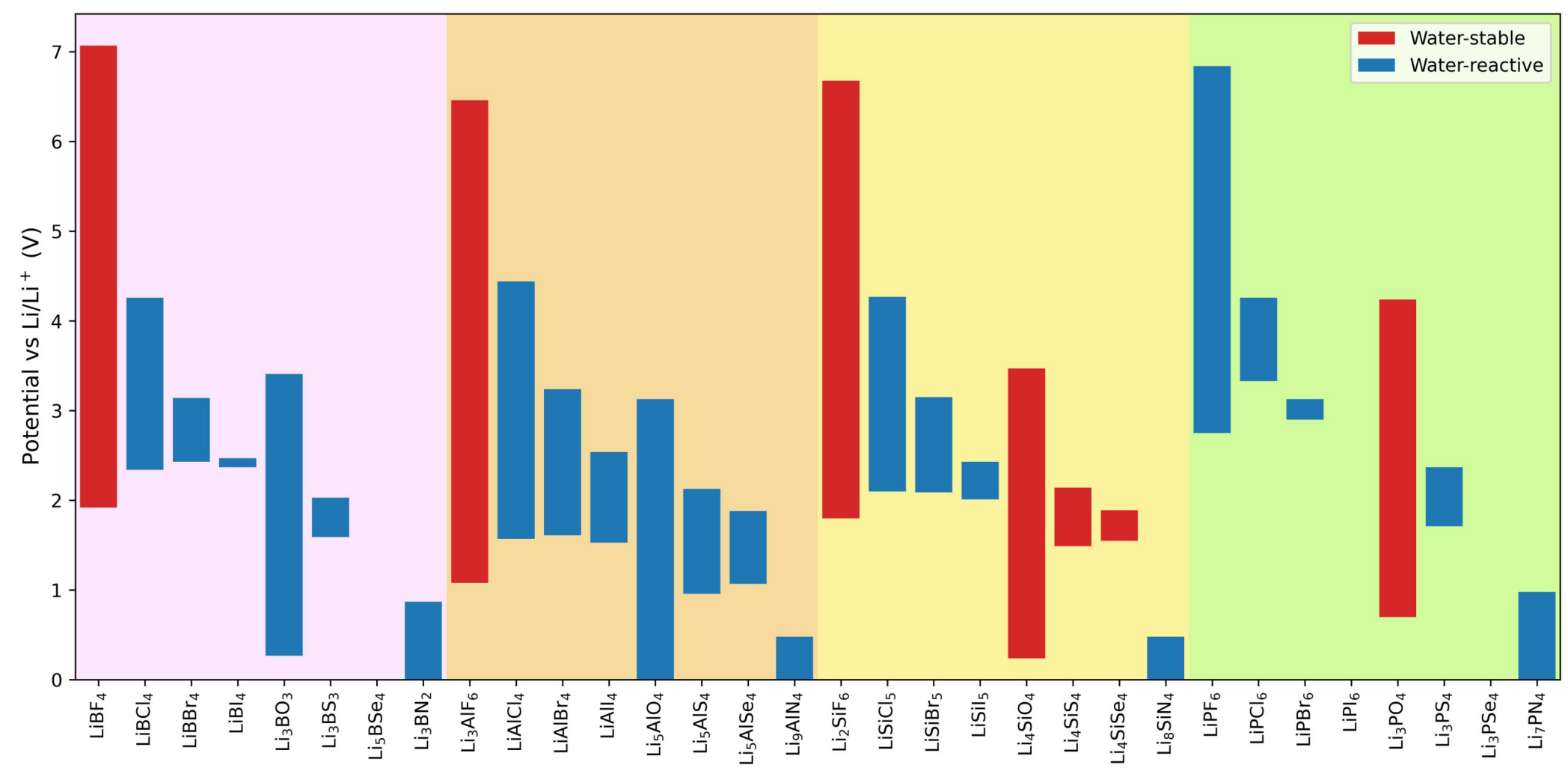

**Figure 4. The electrochemical window and hydrolytic stability for amorphous Li-M-X structures**.

### Bulk modulus for amorphous Li-M-X structures

To further investigate the mechanical properties of amorphous Li-M-X materials, we calculated their bulk moduli. For each structural model, we generated 9 configurations with scaled simulation cell volumes ranging from 0.96 to 1.04 times the original volume. The total energy of each configuration was obtained by structural relaxation and fitted to the Murnaghan equation of state to determine the bulk modulus[35], [36]. The volume-energy relationship for amorphous $Li_5AlO_4$ is representatively shown in Supplementary Figure S12. As Figure 5 shows, the results reveal a significant negative correlation between the bulk modulus and lithium-ion transport. This occurs because in amorphous systems with higher bulk modulus, stronger atomic bonding impedes ionic motion and migration. Thus, bulk modulus can act as a predictive descriptor for ionic transport in amorphous Li-M-X materials. This finding provides scope focused on ternary systems in current work. Whether the correlation holds in doped or multicomponent systems will be explored in subsequent studies.

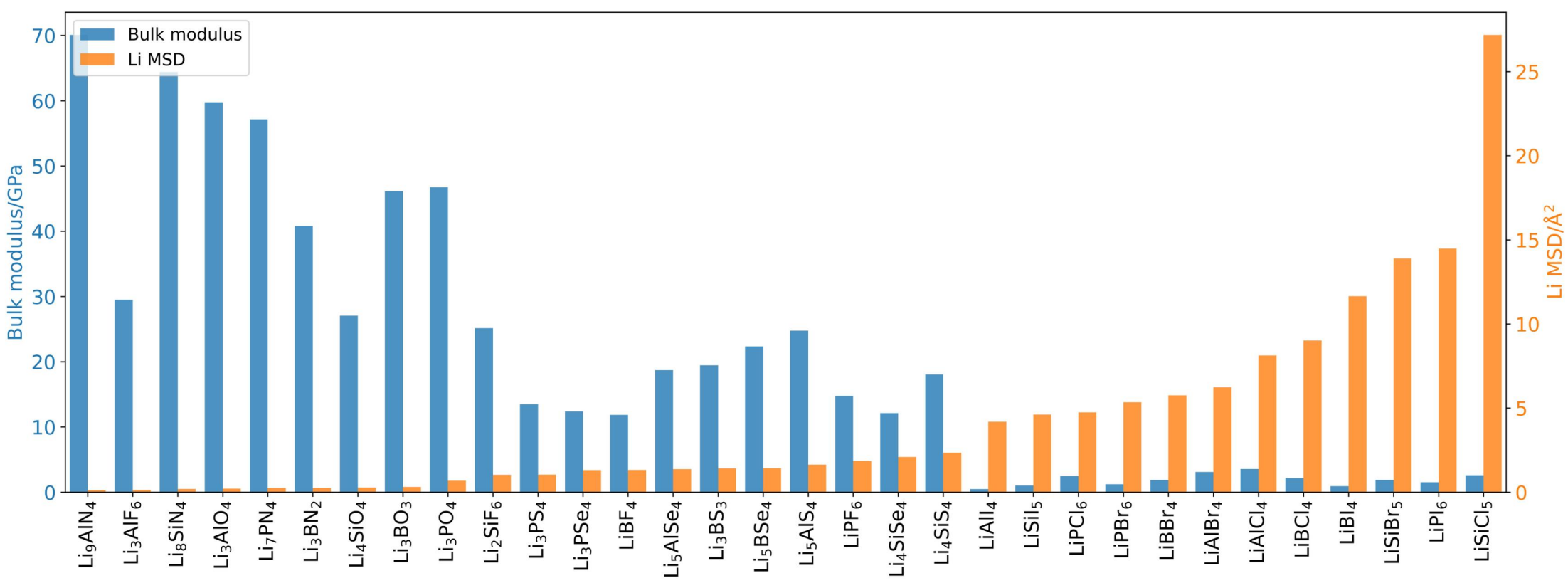

**Figure 5. The correlation between bulk modulus and Li MSD for amorphous Li-M-X structures.**

### Design diagram for amorphous SSEs

It is evident that the type of X anions plays a decisive role in governing ionic migration and related properties. Based on the element type and concentration of X anions, we have constructed a design diagram for amorphous SSEs (Figure 6). In Figure 6, $n$ refers to the coordination number between M cations and X anions in a saturated $MX_n$ polyanionic group, and $c$ refers to the actual atomic ratio of X/M in amorphous $Li_{cx-m}M^{m+}[X^{x-}]_c$.

Here, we will discuss the situations for Li-M-$X^-$, Li-M-$X^{2-}$ and Li-M-$X^{3-}$ respectively, according to their different $Li^+$ ionic transport properties. First, for Li-M-$X^-$ structures, if M cations and isolated $X^-$ anions experience long-distance migration which may decrease the glass forming ability, one can increase the glass forming ability by adding anions with more negative charge (e.g., $O^{2-}$ or $N^{3-}$), stabilize the proportion of $c < n$ to form Type 4 structures with M-X-M chains and reduce the number of isolated $X^-$ anions, such as amorphous $Li_{1.0}AlO_{0.75}Cl_{2.5}$[21]. However, when $c = n$, in Type 2 structures like amorphous $LiBF_4$ where there are limited movement of M elements and a scarcity of isolated $X^-$ anions, the formation of extended chain structures becomes unnecessary. In the above two situations ($c < n$ or $c = n$), the $X^-$ anions rotate around M cations vigorously, forming the paddle-wheel mechanism. Besides, strategic element substitution can achieve $c = n$ to eliminate isolated $X^-$ anions, hence enhance the glass forming ability. An example is the partial replacement of $Cl^-$ anions with $Se^{2-}$ anions in amorphous $LiSiCl_5$ (Type 1 structure), which obtain amorphous $LiSiCl_3Se$ structure (Type 2 structure), eliminating isolated $Cl^-$ anions and remaining high ionic conductivity of 7.12 mS/cm at 300 K simultaneously, as presented in Supplementary Figure S13. However, for amorphous Li-M-$X^-$ structures with $c > n$, mostly Type 1 structures, a considerable portion of isolated $X^-$ anions in structures exhibit long-distance migrations. Hence preparing these structures into crystals, such as crystal $Li_6PS_5Cl$, is more conducive to introducing the structural frustration and generating isolated-anion mechanism, thereby promoting lithium ion migration[31]. Second, for Li-M-$X^{2-}$ structures, when $c \leq n$, with the lithium-ion concentration being sufficiently high, amorphous phase becomes advantageous for systems such as amorphous $Li_3PS_4$ (Type 3 structure). In these materials, $Li^+$ ions' cooperative migration mechanism driven by the vibrational dynamics of $X^{2-}$ anions significantly enhances lithium-ion conduction. When $c > n$, Li-M-$X^{2-}$ materials can be made as amorphous form with cooperative migration mechanism for $Li^+$ ions, such as amorphous $Li_7PS_6$ (Type 1 structure) shown in Supplementary Figure S14, or crystalline form with isolated-anion mechanism, such as $Li_8TiS_6$[31], to gain high ionic conductivity. Finally, for Li-M-$X^{3-}$ structures (e.g. nitrides), $X^{3-}$ can serve as the bridge connecting $MX_a$ polyanionic groups to enhance the glass forming ability of amorphous structures, such as amorphous $Li_{1.3}ZrN_{0.4}Cl_{4.1}$ (Type 4 structure)[20], and they can also be made as crystals with isolated anions, such as crystal $Li_7N_2I$[31].

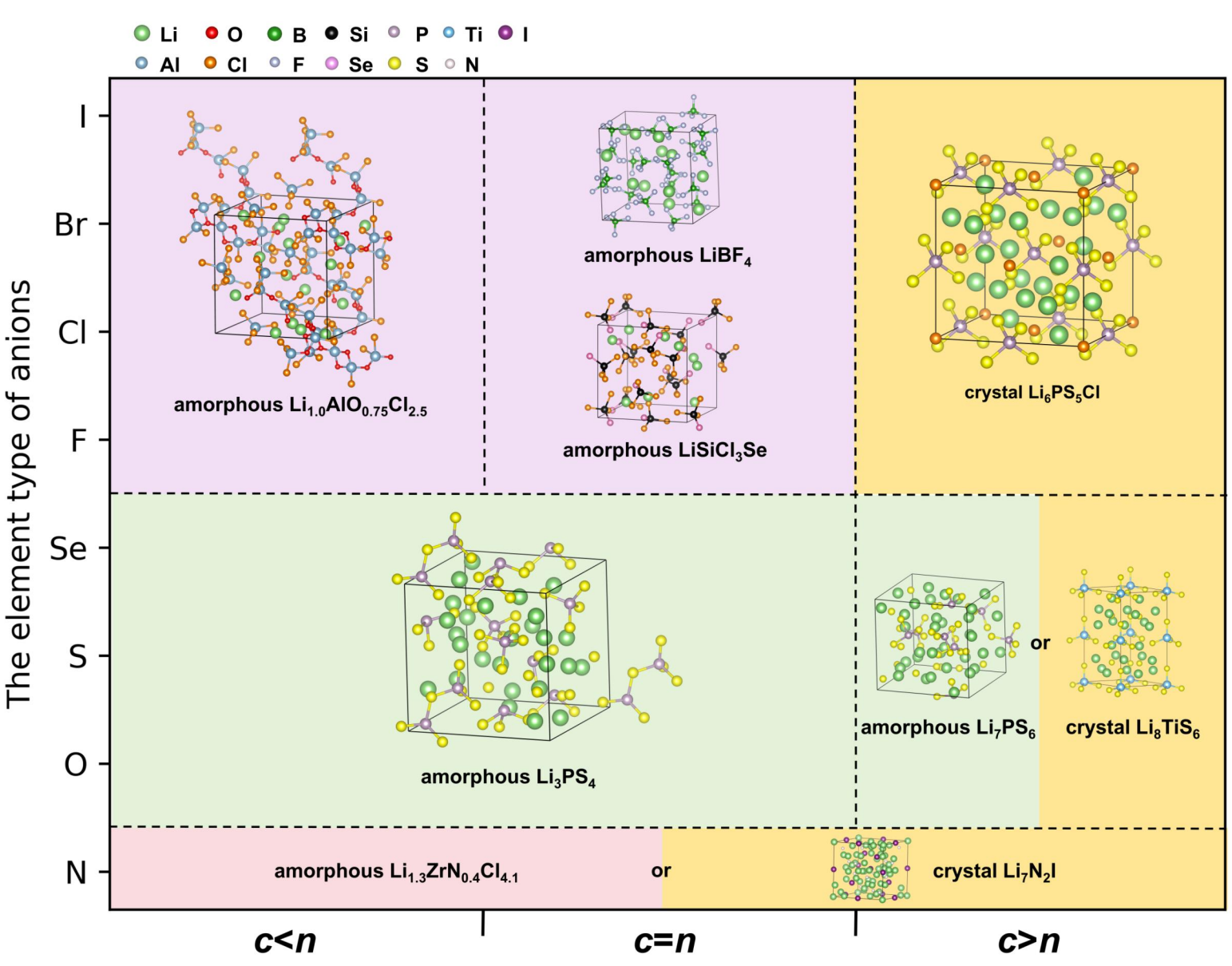

**Figure 6. Design diagram for amorphous SSEs.**

## Conclusions

In this study, by conducting simulations for 32 types of amorphous Li-M-X materials with different M-X pairings, we uncover the universal structural characteristic, ion transport mechanism, electrochemical/moisture stability and mechanical properties for amorphous Li-M-X materials, highlighting the critical role of M-X pairing in determining their properties. By the net charge of $MX_n$ polyanionic groups, we can deduce their propensity for M-X-M chain formation and one can strategically enhance the glass forming ability of amorphous systems. In addition to proposing three $Li^+$ ion transport mechanisms governed by anion-cation coupling (paddle-wheel, cooperative migration and isolated-anion mechanisms) for amorphous SSEs, we extract the conditions triggering anion-cation dynamics, in which atomic disorder and local volume are regarded as two critical factors. Amorphous materials exhibit significantly greater free volume and higher atomic disorder, making lithium-ion transport through dynamic Li-X interactions a more common mechanism compared to their crystalline form. It is also noteworthy that the electrochemical and hydrolytic stability of amorphous SSEs exhibit element-dependent trends similar to those observed in crystalline materials. A strong negative correlation between bulk modulus and ionic conductivity for amorphous Li-M-X materials is identified, and this relationship suggests that computational screening of bulk moduli could serve as a descriptor for identifying potential amorphous ionic conductors among Li-M-X ternary systems. Based on our findings, we establish a comprehensive design diagram and validate it with representative examples, providing a systematic framework for tuning high-performance amorphous SSEs.

While this study establishes a computational framework and design principles for amorphous SSEs, experimental validation is strongly encouraged to test the predictions, particularly for the promising compositions identified.

## Methods

**Density functional theory computation:** All structural relaxations were performed using the Vienna Ab initio Simulation Package (VASP)[37], based on density functional theory (DFT). Calculations employed unit-cells (refer to Supplementary Table S3 for the atom count per cell), the Perdew–Burke–Ernzerhof (PBE)[38] generalized gradient approximation (GGA) functional, and the projector-augmented-wave (PAW) approach. Wavefunction and charge density cutoffs were set to 520 eV and 780 eV, respectively. During optimization, both ionic positions and cell parameters were relaxed, with energy and force convergence criteria of $10^{-5}$ eV and 0.01 eV/Å, respectively. Thermodynamic stability was assessed using the energy above hull ($E_{hull}$), derived from phase diagrams within the Materials Project (MP) database[39], [40]. The electrochemical windows were obtained using GrandPotentialPhaseDiagram package in pymatgen Python package[41] and the reactivity with water was obtained by the ReactionDiagram in pymatgen Python package[41]. We constructed 9 volume-scaled configurations for each structural model, with cell volumes varying from 0.96 to 1.04 times the original size. The structures were then relaxed with ISIF = 4, and subsequently the bulk modulus was obtained by fitting the calculated total energies to the Murnaghan equation of state.

**Ab-initio molecular dynamics simulation:** To investigate the structural and ionic transport properties of amorphous Li-M-X systems (M = B, Al, Si or P; X = F, Cl, Br, I, O, S, Se or N) (Supplementary Table S3 shows the atom number for each simulation cell), we conducted AIMD simulations using lattice parameters near or beyond 10 Å, utilizing nonspin-polarized DFT calculations with a Γ-centered k point grid. The AIMD calculations was performed (time step = 2 fs) in a Nose thermostat[42] with Vienna Ab initio Simulation Package (VASP)[37], using Perdew–Burke–Ernzerhof (PBE)[38], Generalized Gradient Approximation (GGA), employing the projector-augmented-wave (PAW) method. The cutoffs are 520 eV and 780 eV for the wavefunction and the density. All structural and diffusion analysis were performed using the pymatgen and pymatgen-diffusion Python packages.[41], [43] The atomic structures were visualized and analyzed using the VESTA software package.[44] The power spectrum, for $Li^+$ and $Cl^-$ in crystal and amorphous $LiAlCl_4$, was calculated via the Fourier transform of the velocity autocorrelation function derived from AIMD simulations using the same method and formula in Reference [23].

**MLIP-based molecular dynamics simulation:** MLIP models were trained by software NequIP[29] and MD simulations of amorphous $LiBF_4$, $Li_4SiSe_4$ and $LiSiCl_3Se$ structures were conducted by Large-scale Atomic/Molecular Massively Parallel Simulator (LAMMPS)[45] in supercells (Supplementary Table S4 shows the atom number for each simulation cell). The detailed information of the training of MLIP models is presented in Supplementary Note S2. MLIP-based MD simulations were performed with constant volume and temperature (NVT) conditions. The MLIP-based MD simulation process includes 3,000,000 steps at 300 K (time step = 1 fs). The last 2000 ps are used for MSD and $Li^+$ ionic conductivity calculations. The open visualization tool (OVITO)[46] was used to visualize the structures from LAMMPS. The van Hove correlation functions[32] for $Li^+$ ions were calculated from the MLIP-based MD simulations of amorphous $LiBF_4$ and $Li_4SiSe_4$ by the same method used in Reference [47].

**Data availability**

The authors declare that all data supporting the major findings are available in the main text or the Supplementary Information. Source data are provided with this paper.

[1] Zhao, Q., Stalin, S., Zhao, C. Z. & Archer, L. A. Designing solid-state electrolytes for safe, energy-dense batteries. *Nat. Rev. Mater.* **5**, 229-252 (2020).

[2] Luo, J. Y. & Xia, Y. Y. Aqueous lithium-ion battery $LiTi_2(PO_4)_3/LiMn_2O_4$ with high power and energy densities as well as superior cycling stability. *Adv. Funct. Mater.* **17**, 3877-3884 (2007).

[3] Yu, C., Zhao, F. P., Luo, J., Zhang, L. & Sun, X. L. Recent development of lithium argyrodite solid-state electrolytes for solid-state batteries: Synthesis, structure, stability and dynamics. *Nano Energy.* **83**, 105858 (2021).

[4] Adeli, P. et al. Boosting solid-state diffusivity and conductivity in lithium superionic argyrodites by halide substitution. *Angew. Chem. Int. Ed.* **58**, 8681-8686 (2019).

[5] Wu, J. F. et al. Gallium-doped $Li_7La_3Zr_2O_{12}$ garnet-type electrolytes with high lithium-ion conductivity. ACS *Appl. Mater. Interfaces.* **9**, 1542-1552 (2017).

[6] Bernuy-Lopez, C. et al. Atmosphere controlled processing of Ga-substituted garnets for high Li-ion conductivity ceramics. *Chem. Mater.* **26**, 3610-3617 (2014).

[7] Kamaya, N. et al. A lithium superionic conductor. *Nat. Mater.* **10**, 682-686 (2011).

[8] Kato, Y. et al. High-power all-solid-state batteries using sulfide superionic conductors. *Nat. Energy.* **1**, 16030 (2016).

[9] Li, Y. X. *et al.* A lithium superionic conductor for millimeter-thick battery electrode. *Science* **381**, 50-53 (2023).

[10] Aimi, A., Onodera, H., Shimonishi, Y., Fujimoto, K. & Yoshida, S. High Li-ion conductivity in pyrochlore-type solid electrolyte $Li_{2-x}La_{(1+x)/3}M_2O_6F$ (M = Nb, Ta). *Chem. Mater.* **36**, 3717-3725 (2024).

[11] Chandra, A., Bhatt, A. & Chandra, A. Ion conduction in superionic glassy electrolytes: An overview. *J Mater Sci Technol.* **29**, 193-208 (2013).

[12] Grady, Z. A., Wilkinson, C. J., Randall, C. A. & Mauro, J. C. Emerging role of non-crystalline electrolytes in solid-state battery research. *Front. Energy Res.* **8**, 218 (2020).

[13] Wang, W. Family traits. *Nat. Mater.* **11**, 275–276 (2012).

[14] Karthikeyan, A. & Rao, K. J. Structure and silver ion transport in AgI-Ag2MoO4 glasses: A molecular dynamics study. *J. Phys. Chem. B* **101**, 3105-3114 (1997).

[15] Liu, Y. B., Madanchi, A., Anker, A. S., Simine, L. & Deringer, V. L. The amorphous state as a frontier in computational materials design. *Nature Reviews Materials* **10**, 228-241 (2025).

[16] Zhang, S. et al. A family of oxychloride amorphous solid electrolytes for long-cycling all-solid-state lithium batteries. *Nat. Commun.* **14**, 3780 (2023).

[17] Hu, L. et al. A cost-effective, ionically conductive and compressible oxychloride solid-state electrolyte for stable all-solid-state lithium-based batteries. *Nat. Commun.* **14**, 3807 (2023).

[18] Dai, T. et al. Inorganic glass electrolytes with polymer-like viscoelasticity. *Nat. Energy.* **8**, 1221-1228 (2023).

[19] Ishiguro, Y., Ueno, K., Nishimura, S., Iida, G. & Igarashib, Y. $TaCl_5$-glassified ultrafast lithium ion-conductive halide electrolytes for high-performance all-solid-state lithium batteries. *Chem. Lett.* **52**, 237-241 (2023).

[20] Wu, T. T. *et al.* Amorphous nitride-chloride solid-state electrolytes for high performance all-solid-state lithium batteries. *Angew. Chem. Int. Ed.* **64**, e202510359 (2025).

[21] Yang, Q. F. et al. Atomic insight into $Li^+$ ion transport in amorphous electrolytes

$Li_xAlO_yCl_{3+x-2y}$ ($0.5 \leq x \leq 1.5$, $0.25 \leq y \leq 0.75$). *J. Mater. Chem. A.* **13**, 2309-2315 (2025).
[22] Lei, M., Li, B., Liu, H. J. & Jiang, D. E. Dynamic monkey bar mechanism of superionic Li-ion transport in $LiTaCl_6$. *Angew. Chem. Int. Ed.* **63**, e202315628 (2024).
[23] Smith, J. G. & Siegel, D. J. Low-temperature paddlewheel effect in glassy solid electrolytes. *Nat. Commun.* **11**, 1483 (2020).
[24] Xu, Z. M. et al. Machine learning molecular dynamics simulation identifying weakly negative effect of polyanion rotation on Li-ion migration. *Npj Comput. Mater.* **9**, 105 (2023).
[25] Jun, K., Lee, B., Kam, R. L. & Ceder, G. The nonexistence of a paddlewheel effect in superionic conductors. *Proc. Natl. Acad. Sci. U. S. A.* **121**, e2316493121 (2024).
[26] Smith, J. G. & Siegel, D. J. A proper definition of the paddlewheel effect affirms its existence. *Proc. Natl. Acad. Sci. U. S. A.* **122**, e2419892122 (2025).
[27] Li, K., Yang, J. T., Zhai, Y. & Li, H. Disentangling cation-polyanion coupling in solid electrolytes: Which anion motion dominates cation transport? arXiv:2501.02440.
[28] Lee, B., Jun, K., Ouyang, B. & Ceder, G. Weak correlation between the polyanion environment and ionic conductivity in amorphous Li-P-S superionic conductors. *Chem. Mater.* **35**, 891-899 (2023).
[29] Batzner, S. et al. E(3)-equivariant graph neural networks for data-efficient and accurate interatomic potentials. *Nat. Commun.* **13**, 2453 (2022).
[30] Dietrich, C. et al. Lithium ion conductivity in $Li_2S$-$P_2S_5$ glasses - building units and local structure evolution during the crystallization of superionic conductors $Li_3PS_4$, $Li_7P_3S_{11}$ and $Li_4P_2S_7$. *J. Mater. Chem. A.* **5**, 18111-18119 (2017).
[31] Yang, Q. F., J. Xu, Y. Q. Wang, X. Fu, R. J. Xiao and H. Li. New fast ion conductors discovered through the structural characteristic involving isolated anions. *Npj Comput. Mater.* **11**, 67 (2025).
[32] Vanhove, L. Correlations in space and time and born approximation scattering in systems of interacting particles. *Physical Review* **95**, 249-262 (1954).
[33] Zhu, Y. Z., He, X. F. & Mo, Y. F. First principles study on electrochemical and chemical stability of solid electrolyte-electrode interfaces in all-solid-state Li-ion batteries. *J. Mater. Chem. A.* **4**, 3253-3266 (2016).
[34] Lin, S. et al. Reclaiming neglected compounds as promising solid state electrolytes by predicting electrochemical stability window with dynamically determined decomposition pathway. *Adv. Energy Mater.* **12**, 2201808 (2022).
[35] Murnaghan, F. D. The compressibility of media under extreme pressures. *Proc. Natl. Acad. Sci. U. S. A.* **30**, 244-247 (1944).
[36] Tanibata, N. et al. Metastable chloride solid electrolyte with high formability for rechargeable all-solid-state lithium metal batteries. *ACS Mater. Lett.* **2**, 880-886 (2020).
[37] Kresse, G. & Furthmuller, J. Efficiency of ab-initio total energy calculations for metals and semiconductors using a plane-wave basis set. *Comput. Mater. Sci.* **6**, 15-50 (1996).
[38] Perdew, J. P., Burke, K. & Ernzerhof, M. Generalized gradient approximation made simple. *Phys Rev Lett.* **77**, 3865-3868 (1996).
[39] Xu, J. et al. New halide-based sodium-ion conductors $Na_3Y_2Cl_9$ inversely designed by building block construction. *ACS Appl. Mater. Interfaces.* **15**, 21086 (2023).
[40] Ong, S. P.,Wang, L., Kang, B. & Ceder, G. Li−Fe−P−$O_2$ Phase diagram from first principles calculations. *Chem. Mater.* **20**, 1798 (2008).
[41] Ong, S. P. et al. Python Materials Genomics (pymatgen): A robust, open-source python library for materials analysis. *Comput. Mater. Sci.* **68**, 314-319 (2013).
[42] Nose, S. Constant temperature molecular-dynamics methods. *Theor. Phys. Suppl.* **103**, 1-46 (1991).
[43] Deng, Z., Zhu, Z. Y., Chu, I. H. & Ong, S. P. Data-driven first-principles methods for the study and design of alkali superionic conductors. *Chem. Mater.* **29**, 281-288 (2017).
[44] Momma, K. & Izumi, F. VESTA 3 for three-dimensional visualization of crystal, volumetric and morphology data. *J Appl Crystallogr.* **44**, 1272-1276 (2011).
[45] Plimpton, S. Fast parallel algorithms for short-range molecular-dynamics. *J. Comput. Phys.* **117**, 1-19 (1995).

[46] Stukowski, A. Visualization and analysis of atomistic simulation data with OVITO-the Open Visualization Tool. *Model Simul Mat Sci Eng* **18**, 015012 (2010).
[47] He, X. F., Zhu, Y. Z. & Mo, Y. F. Origin of fast ion diffusion in super-ionic conductors. *Nature Communications* **8**, 15893 (2017).

**Acknowledgements:**
We thank the computing resources provided by ORISE Supercomputer, and the National Supercomputer Center in Tianjin.

**Funding:**
This work was financially supported by Beijing Natural Science Foundation (Z250005) and the Strategic Priority Research Program of Chinese Academy of Sciences (grant no. XDB0500201).

**Author contributions:**
R. X. supervised and provided guidance for the entire project. Q. Y. completed the calculation and analysis of the data, as well as the writing of the article. X. F., X. G., J. L., L. W., Y. H. and H. L. provided their own suggestions for the project. All authors contributed to the discussions and revisions of the manuscript.

**Competing interests:**
The authors declare no competing interests.

**Supplementary information:**
Supplementary Figures S1 to S14
Supplementary Tables S1 to S4
Supplementary Notes S1 and S2

# Supporting Information for

## Design principles for amorphous solid-state electrolytes

Qifan Yang[1, 2], Xiao Fu[1, 2], Xuhe Gong[1, 4], Jingchen Lian[1, 3], Liqi Wang[1, 3], Ruijuan Xiao[1, 2, 3, *],

Yong-Sheng Hu[1, 2, 3, *] and Hong Li[1, 2, 3, *]

[1] Beijing National Laboratory for Condensed Matter Physics, Institute of Physics, Chinese Academy of Sciences, Beijing 100190, China

[2] Center of Materials Science and Optoelectronics Engineering, University of Chinese Academy of Sciences, Beijing 100049, China

[3] School of Physical Sciences, University of Chinese Academy of Sciences, Beijing 100049, China

[4] School of Materials Science and Engineering, Key Laboratory of Aerospace Materials and Performance (Ministry of Education), Beihang University, Beijing 100191, China

E-mail: rjxiao@iphy.ac.cn, yshu@iphy.ac.cn, hli@iphy.ac.cn

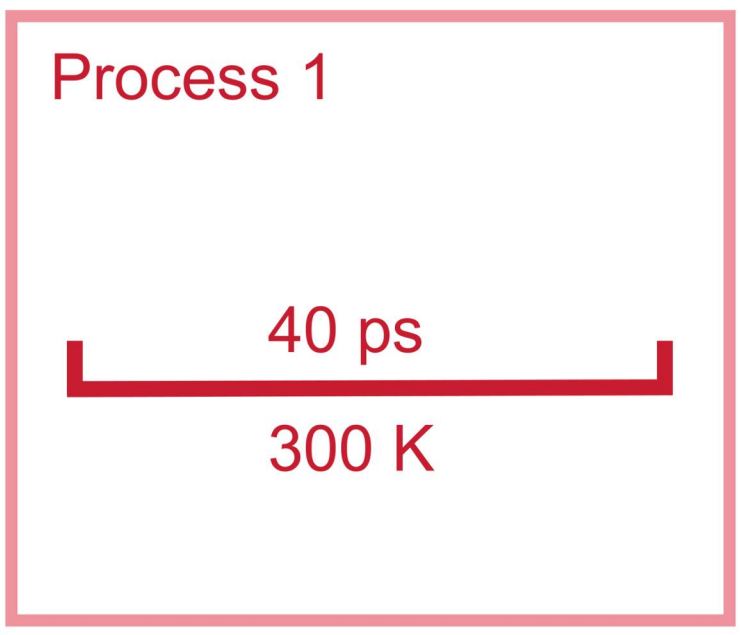

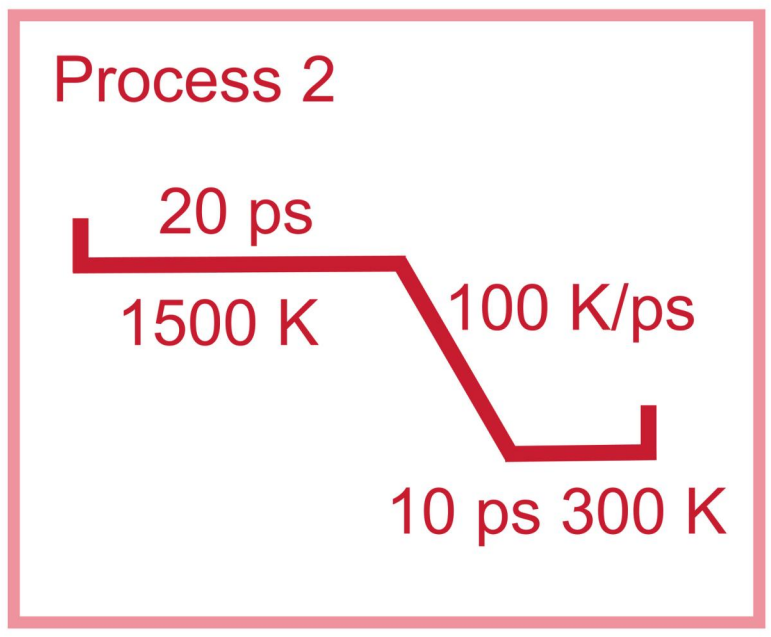

**Supplementary Figure S1. The Process 1 and Process 2 for initial disordered Li-M-X structures conducted by ab-initio molecular dynamics (AIMD) simulations.** The similarity of generated amorphous structures in Process 1 and Process 2 is evaluated by assessing the consistency of their free energy and radial distribution function (RDF) curves.

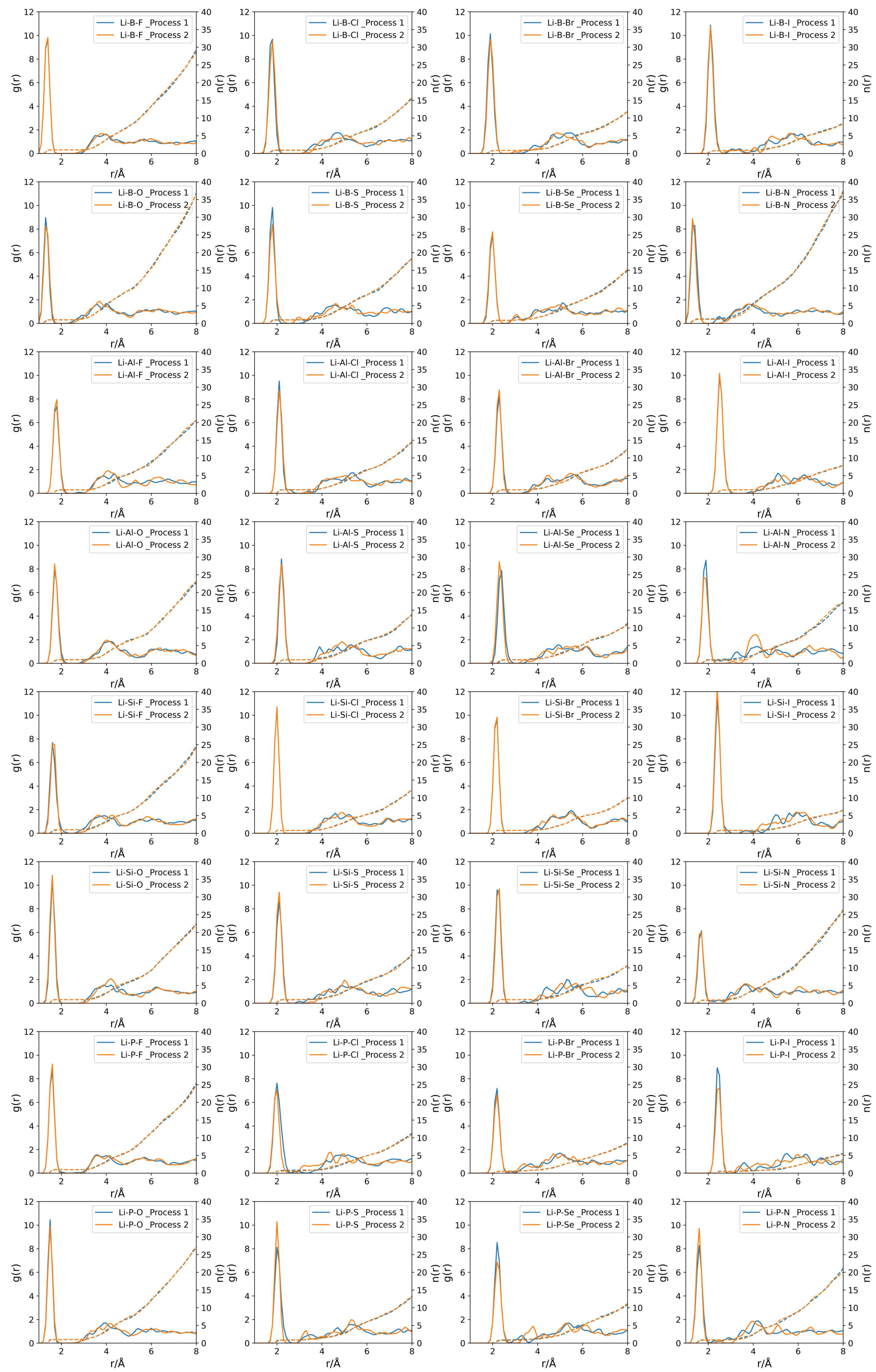

**Supplementary Figure S2. The RDF curves and coordination number curves of the amorphous Li-M-X structures.** The results show that the RDF and coordination number curves of the structures generated by Process 1 and 2 are very similar for all 32 Li-M-X amorphous samples. Therefore, we primarily distinguish between the structures produced by the two processes based on the energy differences presented in Supplementary Table S1.

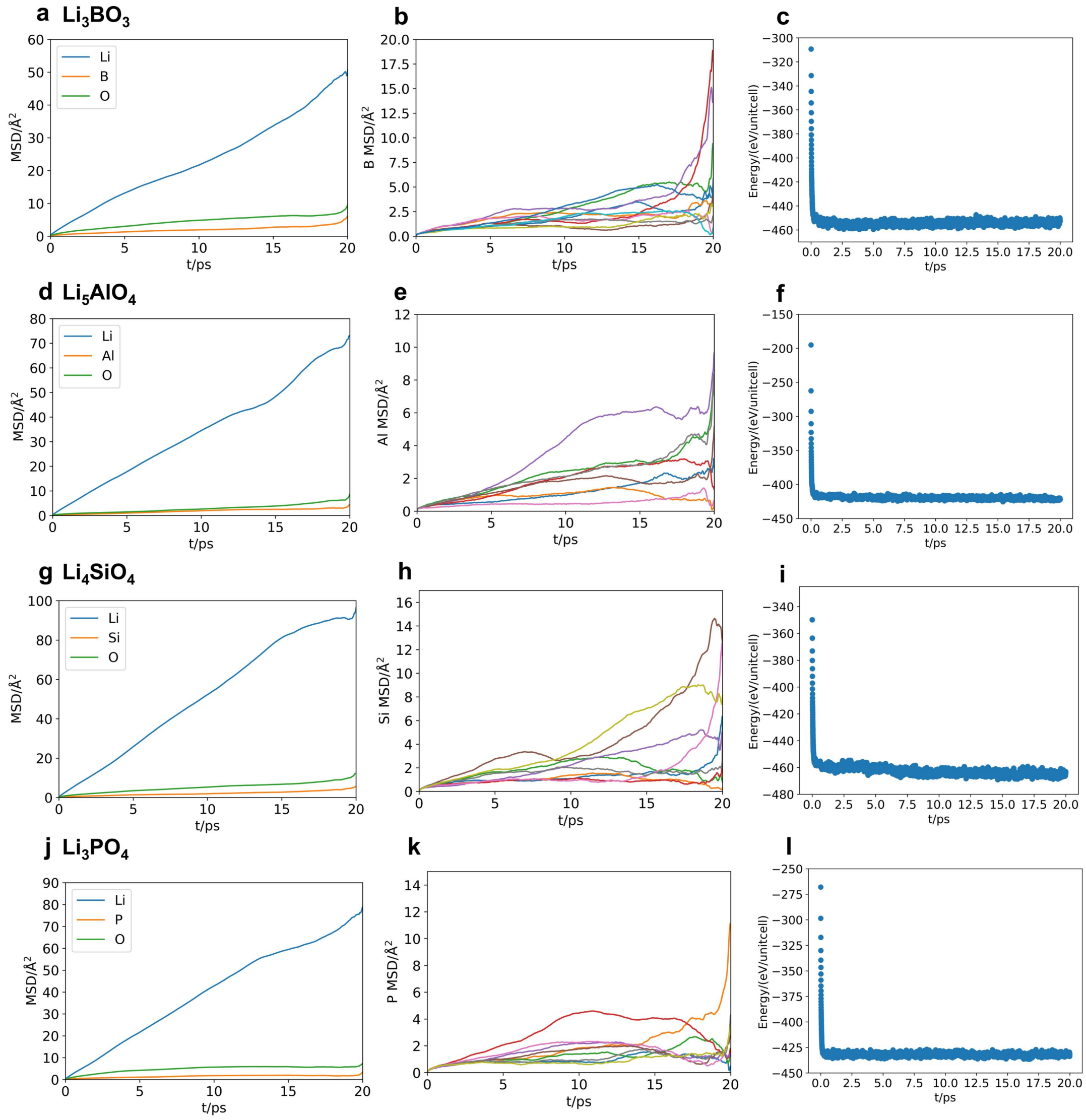

**Supplementary Figure S3. Evidence that amorphous Li-M-X oxides underwent melting and achieved equilibration at 1500 K in Process 2, including MSDs for each element, MSDs for individual M cations and the variation of energy per unit cell over time of $Li_3BO_3$ (a-c), $Li_5AlO_4$ (d-f), $Li_4SiO_4$ (g-i) and $Li_3PO_4$ (j-l).** First, the MSDs indicate that even the slowest-moving M atoms in the structures have melted. Second, the energy fluctuation curves demonstrate that the structures have fully equilibrated at 1500 K. These evidences prove that in Process 2, with the exception of nitrides, the most refractory structures - oxides - have melted at 1500 K, suggesting that all structures except nitrides have melted at 1500 K.

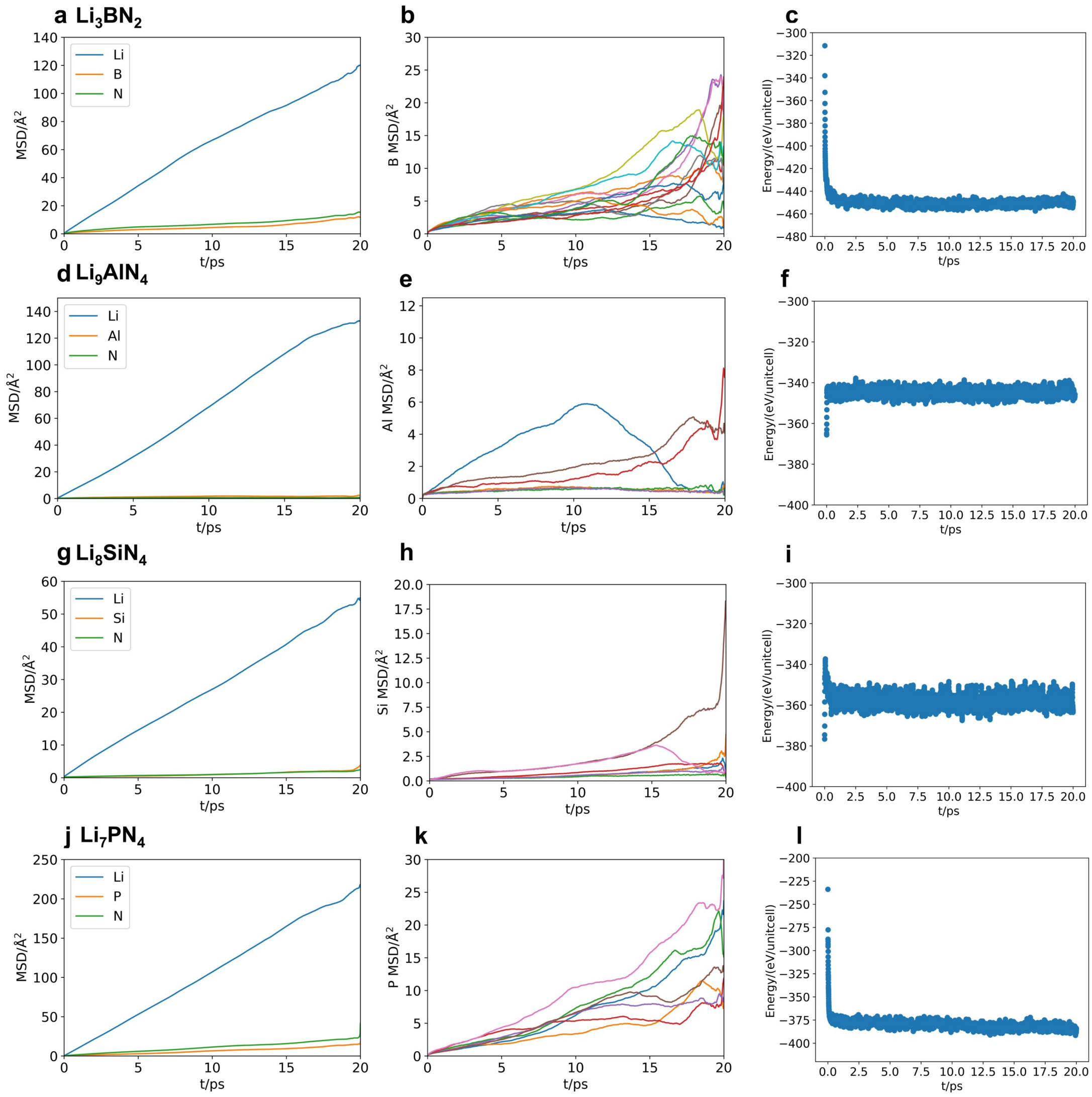

**Supplementary Figure S4. Evidence that amorphous Li-M-X nitrides underwent melting and achieved equilibration the corresponding temperatures (1900 K for $Li_3BN_2$ and $Li_9AlN_4$; 2500 K for $Li_8SiN_4$ and $Li_7PN_4$), including MSDs for each element, MSDs for individual M cations and the variation of energy per unit cell over time of $Li_3BN_2$ (a-c), $Li_9AlN_4$ (d-f), $Li_8SiN_4$ (g-i) and $Li_7PN_4$ (j-l).** First, the MSDs indicate that even the slowest-moving M atoms in the structures have melted. Second, the energy fluctuation curves demonstrate that the structures have fully equilibrated at the corresponding temperatures. These evidences prove that all nitrides have melted at the corresponding temperatures.

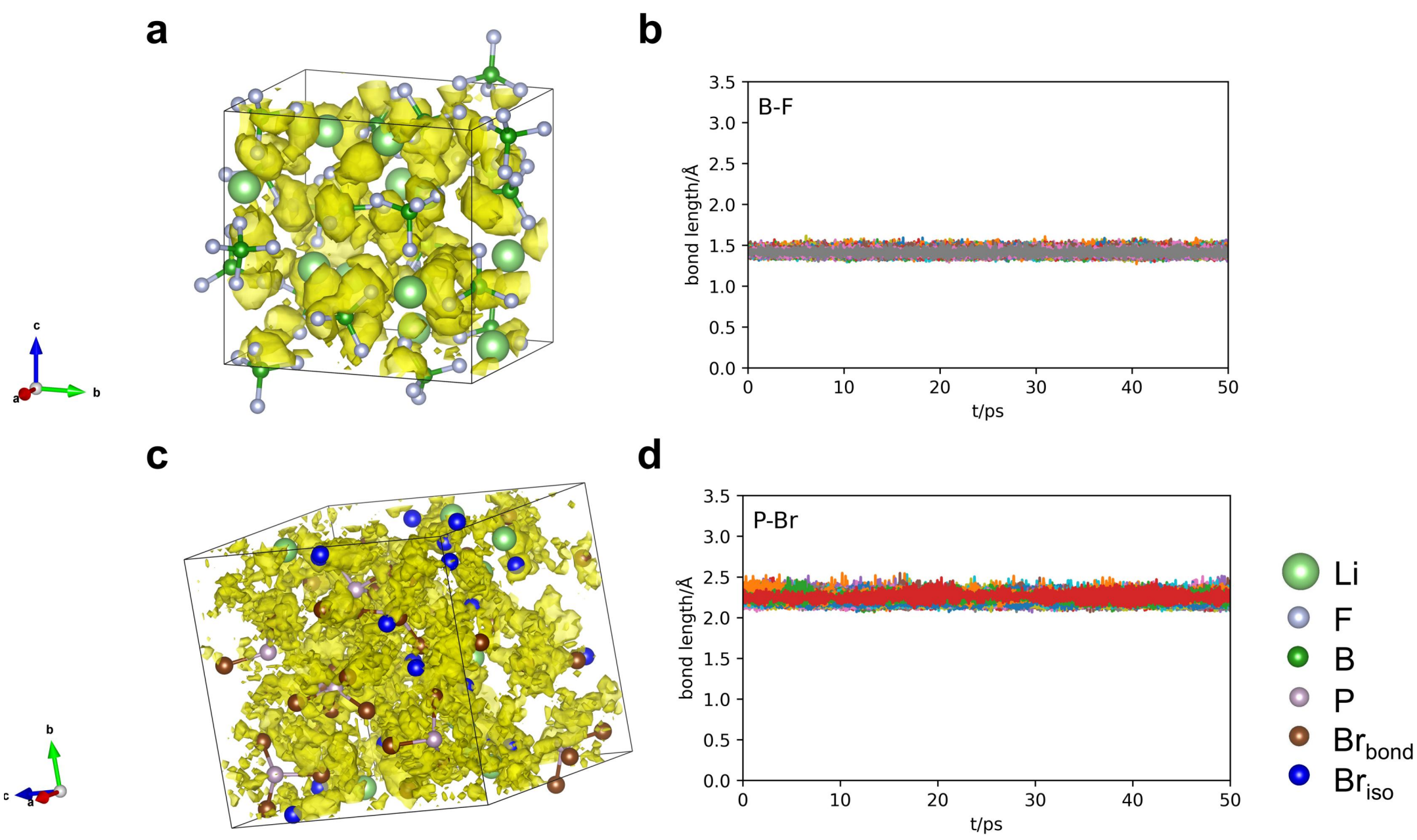

**Supplementary Figure S5. The direct evidence of X anion rotation around M cation within M-X polyanionic groups in amorphous Li-M-X structures.**

**a** The migration path of $F^-$ anions in amorphous $LiBF_4$ structure (the yellow part) during the AIMD simulation at 300 K. **b** The variation of B-F bond length over time during the 300 K AIMD simulation for amorphous $LiBF_4$. **c** The migration path of $Br^-$ anions in amorphous $LiPBr_6$ structure (the yellow part) during the AIMD simulation at 300 K. **d** The variation of P-Br bond length over time during the 300 K AIMD simulation for amorphous $LiPBr_6$.

It can be seen that during the AIMD simulations at 300 K, the movement of X anions in both systems is very intense, yet the B-F or P-Br bond length does not significantly increase, indicating the rotation of $F^-$ and $Br^-$, verifying the paddle-wheel mechanism.

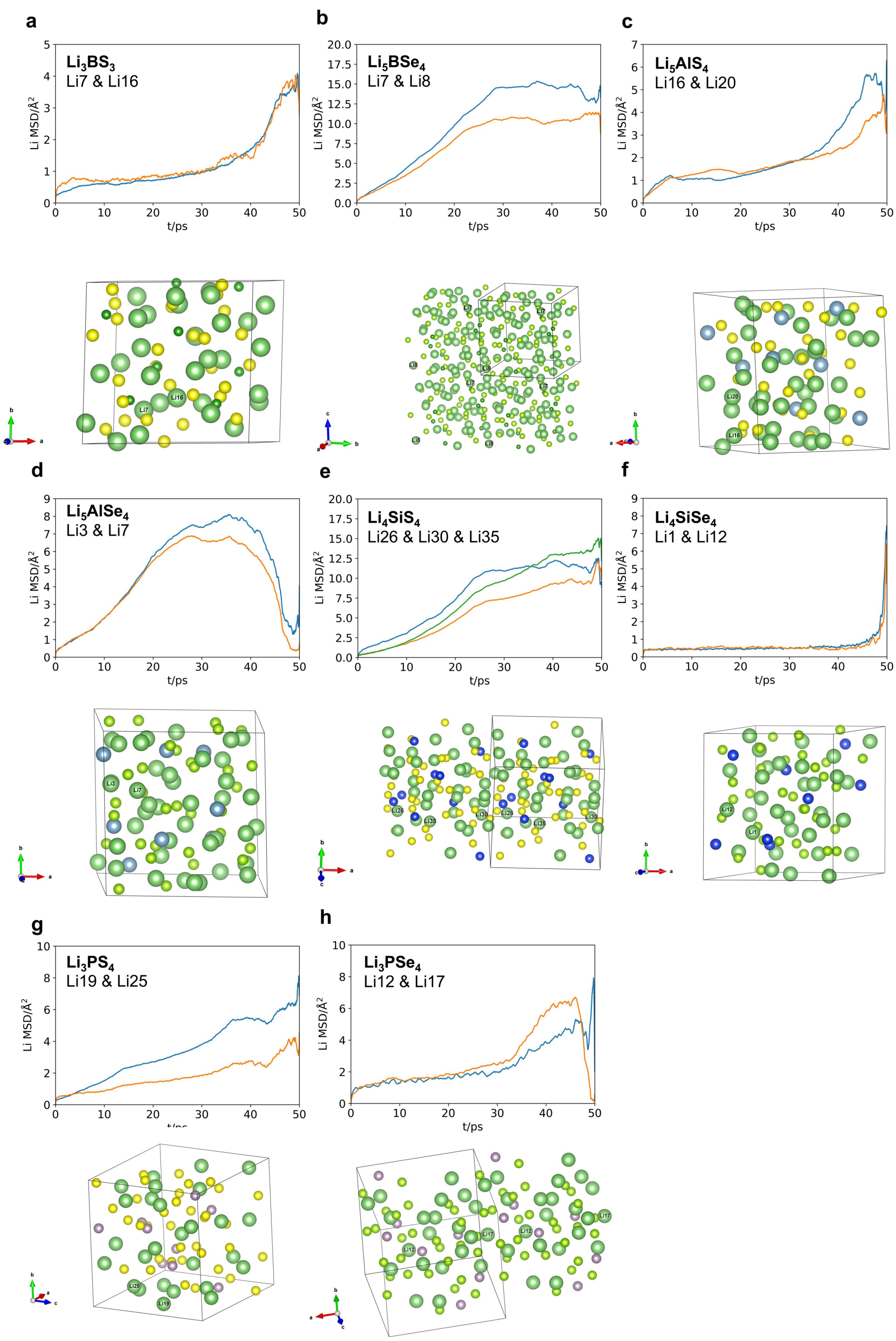

**Supplementary Figure S6. The mean squared displacement (MSD) and position of $Li^+$ ions occurring cooperative migration in amorphous $Li_3BS_3$ (a), $Li_5BSe_4$ (b), $Li_5AlS_4$ (c), $Li_5AlSe_4$ (d), $Li_4SiS_4$ (e), $Li_4SiSe_4$ (f), $Li_3PS_4$ (g) and $Li_3PSe_4$ (h) structures during AIMD simulations at 300 K.**

Similar MSD curves of neighbored $Li^+$ ions in every structure indicate the presence of cooperative migrations for lithium ions.

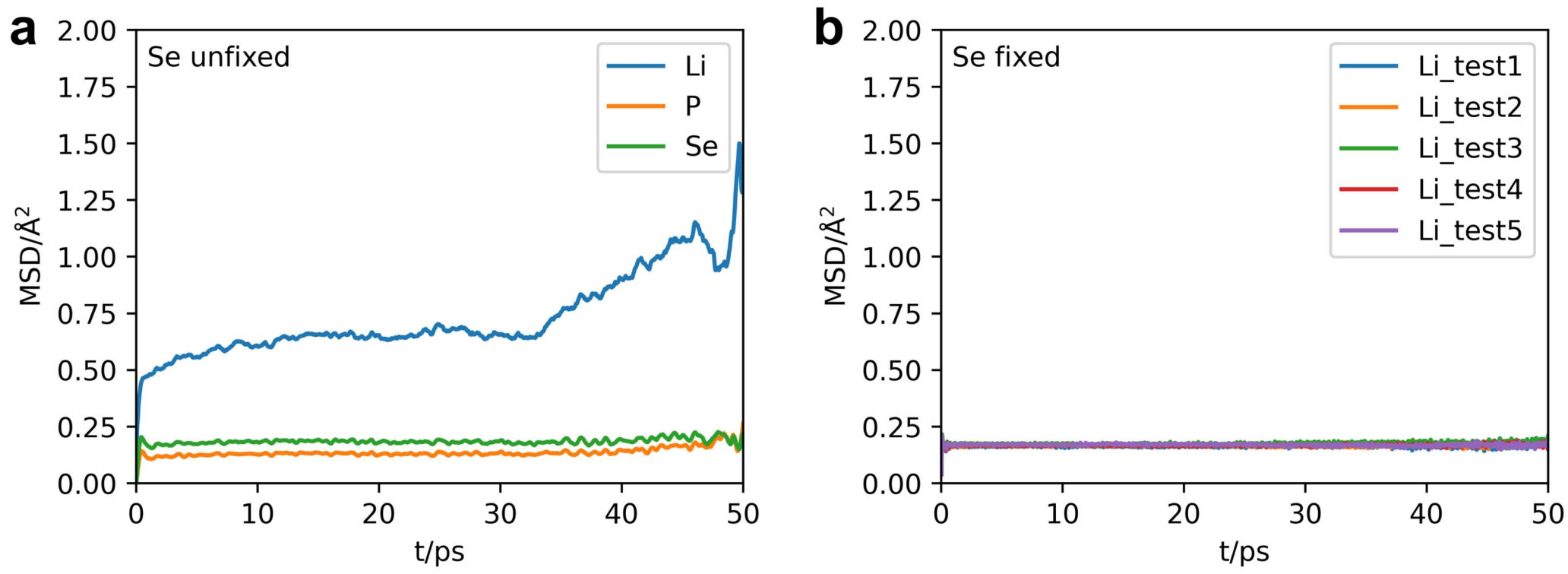

**Supplementary Figure S7. MSDs for amorphous $Li_3PSe_4$ with Se anions unfixed (a) and fixed (b).**

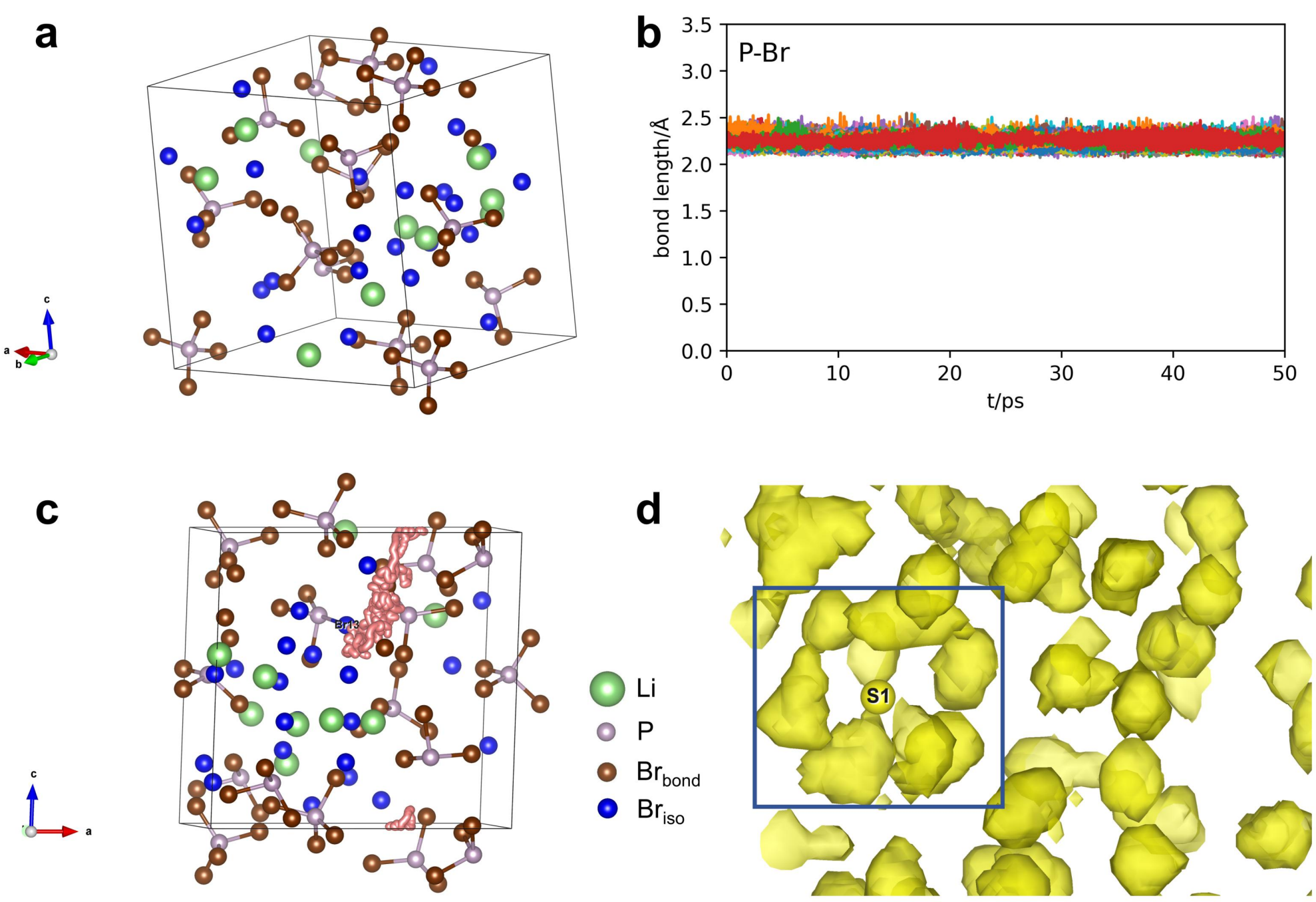

**Supplementary Figure S8. The effect of isolated anions in amorphous $LiPBr_6$ and $Li_4SiS_4$. a** Structure of amorphous $LiPBr_6$. **b** The variation of P-Br bond length over time during the 300 K AIMD simulation for amorphous $LiPBr_6$. **c** The transport path of one isolated $Br^-$ anion during the 300 K AIMD simulation for amorphous $LiPBr_6$. **d** Cage transport of $Li^+$ ions in amorphous $Li_4SiS_4$. The yellow part represents the migration path of lithium ions during AIMD simulations 300 K.

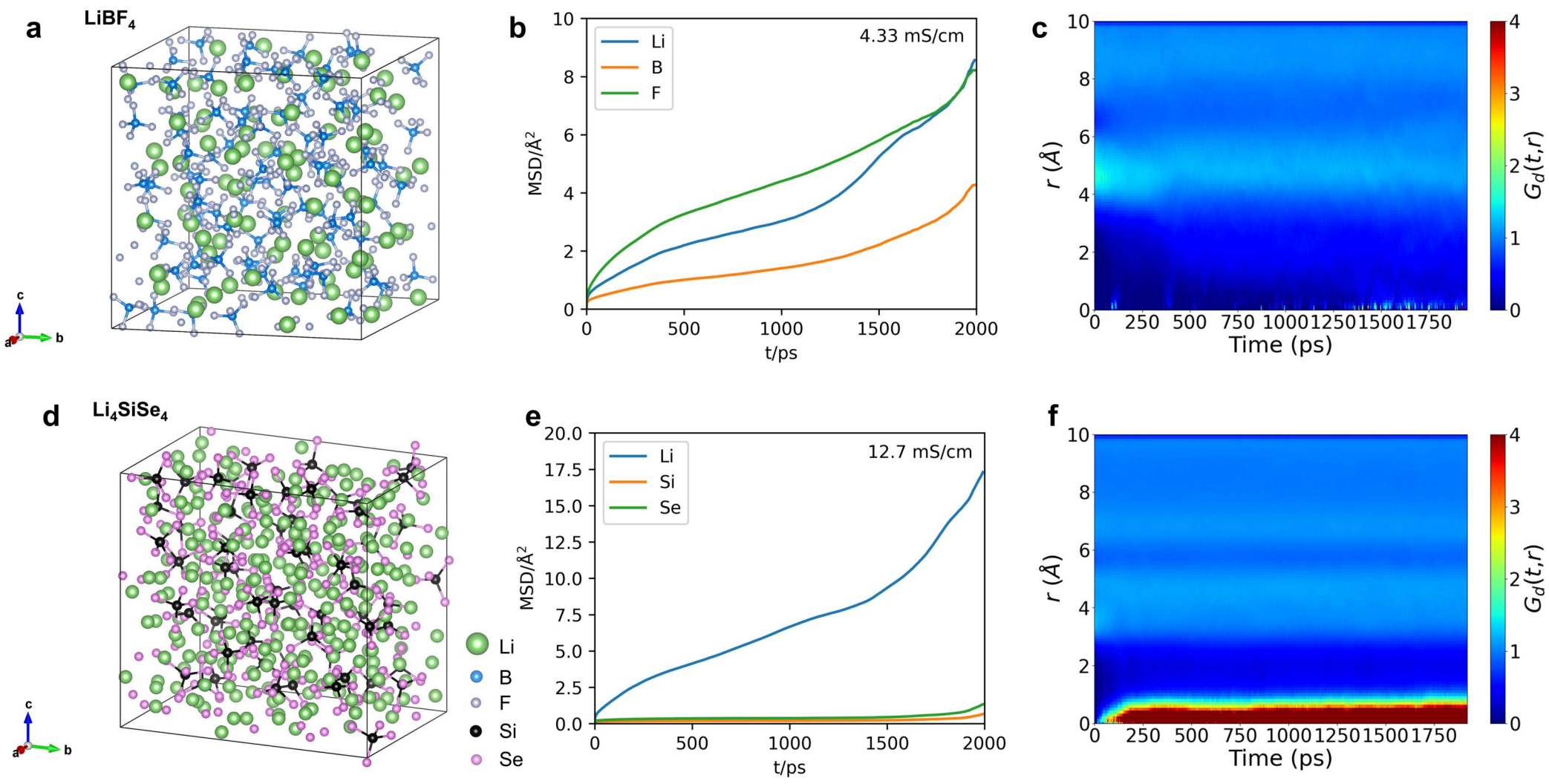

**Supplementary Figure S9. The results of machine-learning interatomic potential based molecular dynamics (MLIP-based MD) simulations for amorphous $LiBF_4$ and $Li_4SiSe_4$.**

**a** Amorphous $LiBF_4$ structure obtained by MLIP-based MD simulations at 300 K. **b** MSDs for each element of amorphous $LiBF_4$ during MLIP-based MD simulations at 300 K. **c** The van Hove correlation functions of $Li^+$ dynamics on distinctive Li ions of amorphous $LiBF_4$ during MLIP-based MD simulations at 300 K. **d** Amorphous $Li_4SiSe_4$ structure obtained by MLIP-based MD simulations at 300 K. **e** MSDs for each element of amorphous $Li_4SiSe_4$ during MLIP-based MD simulations at 300 K. **f** The van Hove correlation functions of $Li^+$ dynamics on distinctive Li ions of amorphous $Li_4SiSe_4$ during MLIP-based MD simulations at 300 K.

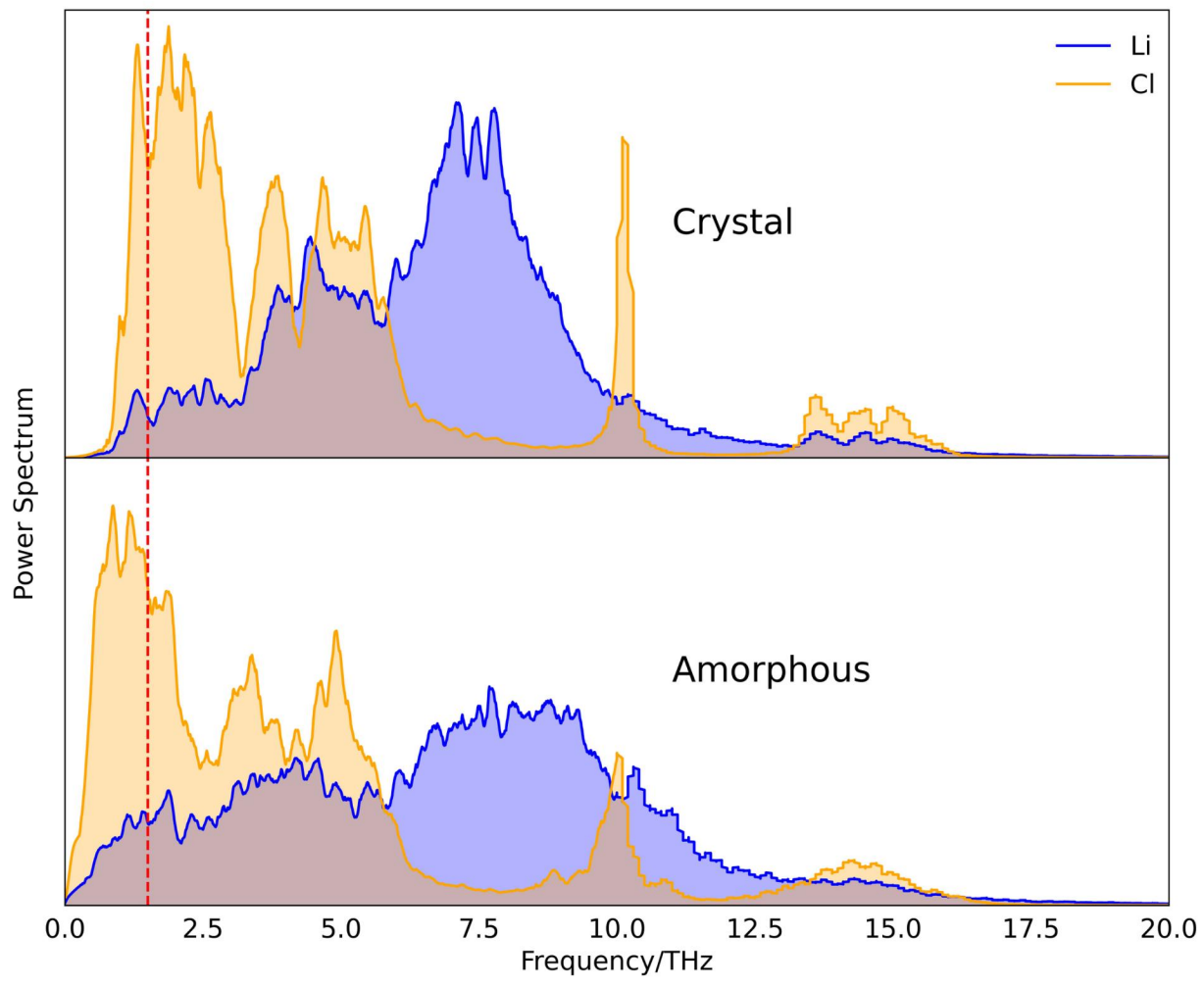

**Supplementary Figure S10. The velocity autocorrelation power spectrum of $Li^+$ and $Cl^-$ in both crystalline and amorphous $LiAlCl_4$.**

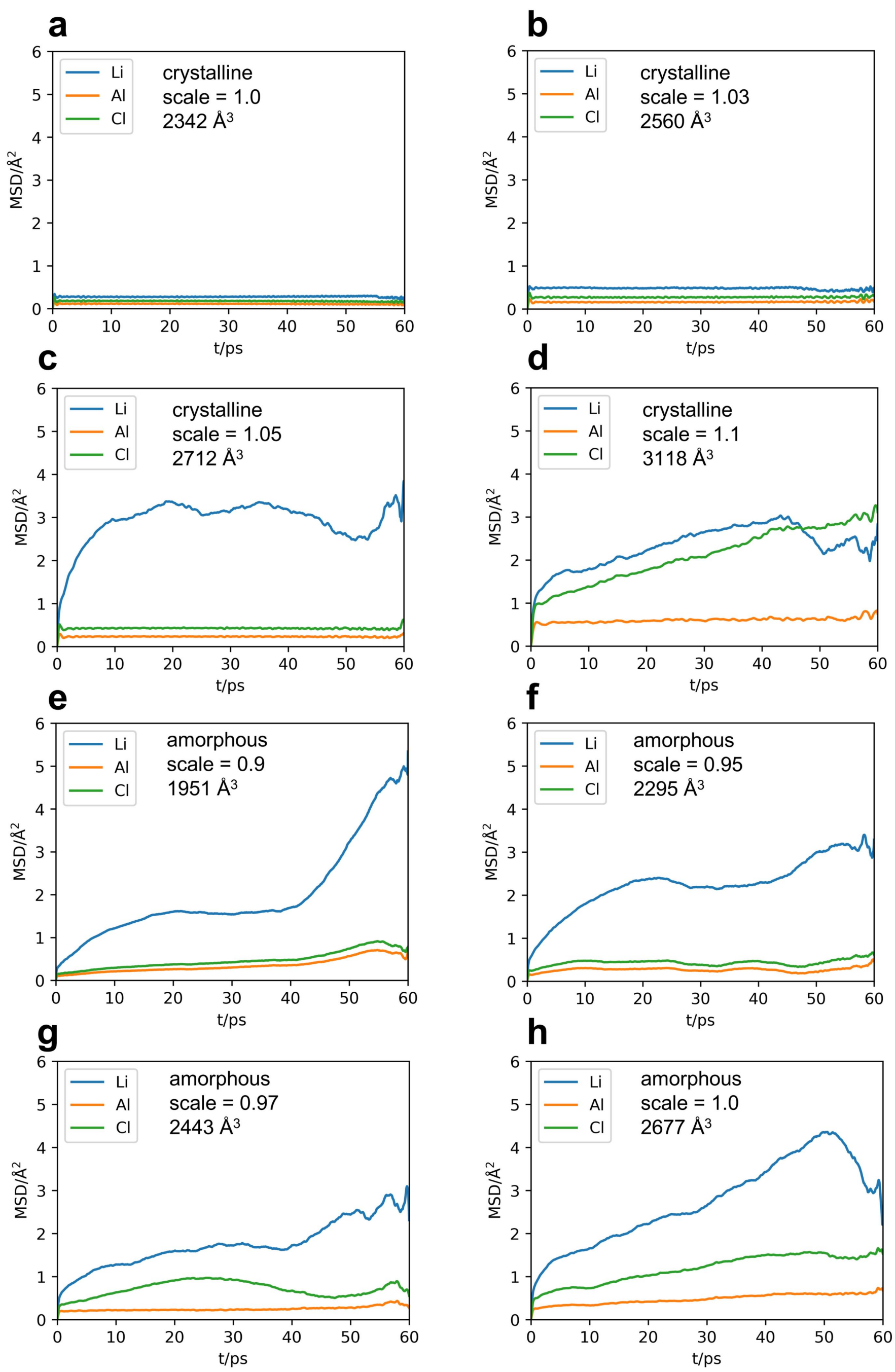

**Supplementary Figure S11. MSDs for each element of crystalline and amorphous $LiAlCl_4$ structures with different volume of unit-cells at 300 K.** **a** presents the baseline crystalline structure, while **b-d** display the same crystalline cell with its expanded to 1.03, 1.05 and 1.10 times the original, respectively; **e-g** depict the amorphous structure with its volume reduced to 0.90, 0.95 and 0.97 of the original, respectively, and **h** shows the amorphous system at its original volume. Each subplot plots MSD for each element during AIMD simulations at 300 K.

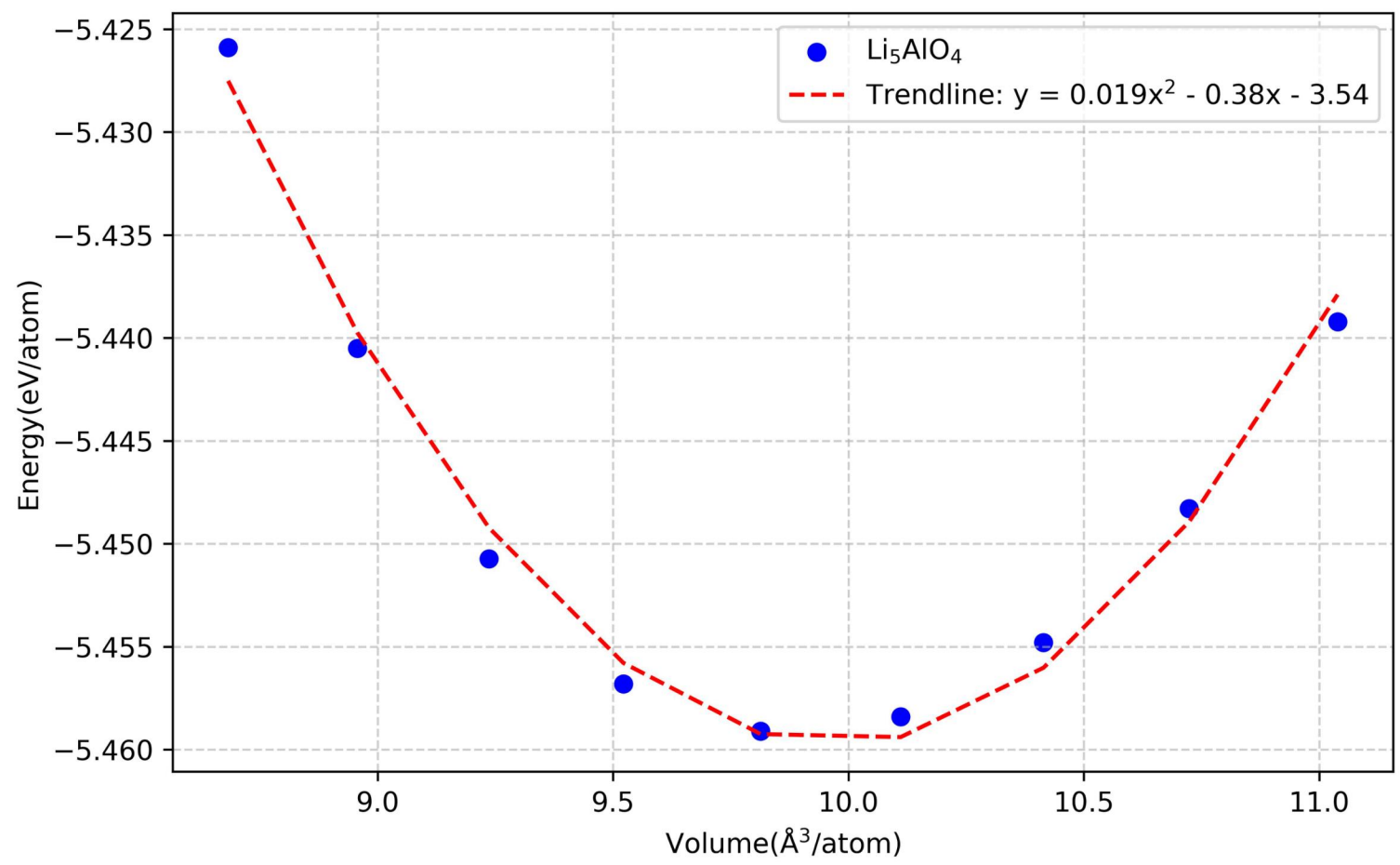

**Supplementary Figure S12. The volume-energy relationship for amorphous $Li_5AlO_4$.**

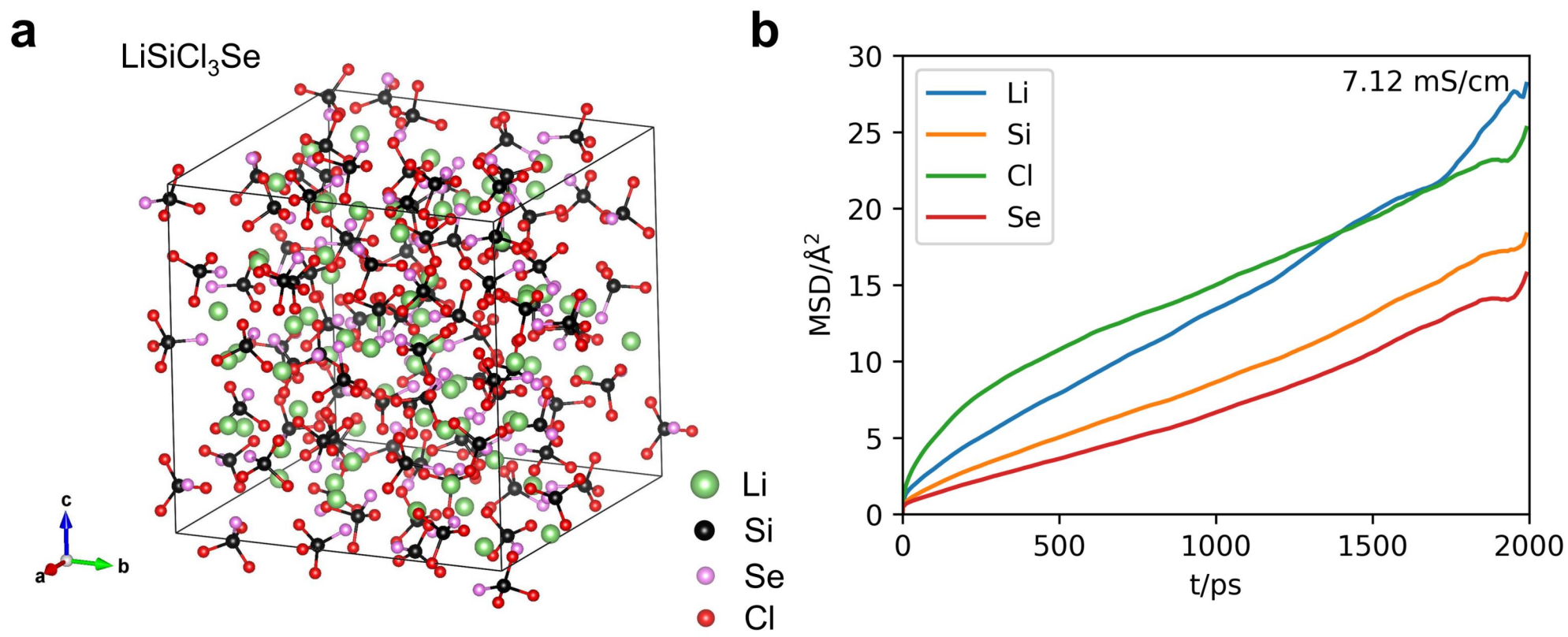

**Supplementary Figure S13. The structure (a) and MSDs for each element at 300 K (b) of amorphous $LiSiCl_3Se$.**

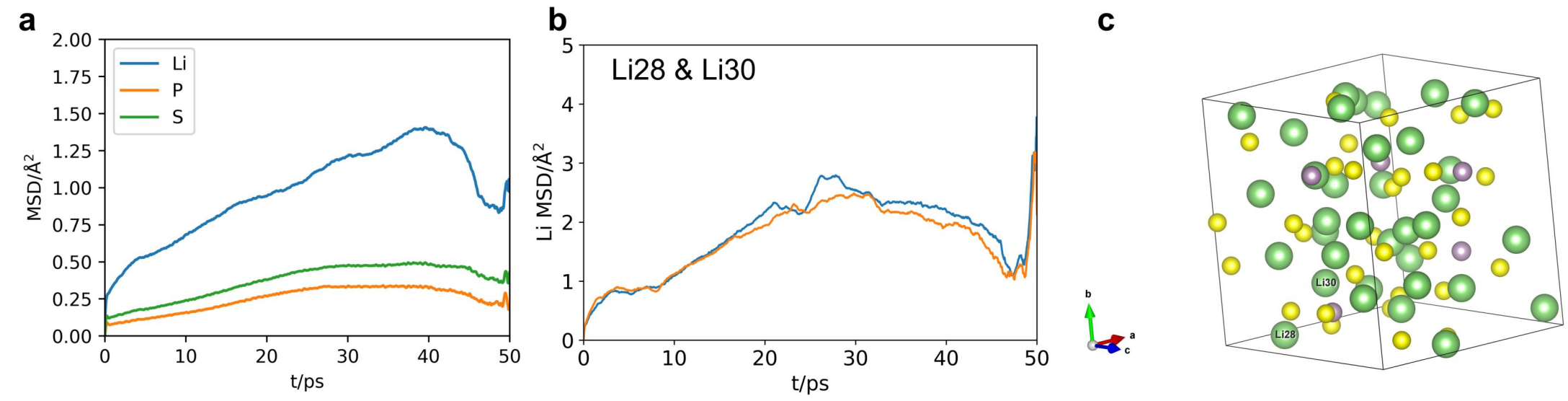

**Supplementary Figure S14. The ionic transport properties of amorphous $Li_7PS_6$ during AIMD simulations at 300 K. a** MSDs for each element of amorphous $Li_7PS_6$ during AIMD simulations at 300 K. **b-c** The MSD and position of $Li^+$ ions occurring cooperative migration in amorphous $Li_7PS_6$ during AIMD simulations at 300 K.

**Supplementary Table S1. Free energies of amorphous Li-M-X structures produced by Process 1 and 2, and their difference.**

| Structure | Free energy of structures produced by Process 1/ (eV/atom) | Free energy of structures produced by Process 2/ (eV/atom) | Energy difference/ (eV/atom) |
|---|---|---|---|
| $LiBF_4$ | -5.56 | -5.56 | 0 |
| $LiBCl_4$ | -3.80 | -3.81 | 0.01 |
| $LiBBr_4$ | -3.28 | -3.29 | 0.01 |
| $LiBI_4$ | -2.79 | -2.78 | -0.01 |
| $Li_3BO_3$ | -6.07 | -6.04 | -0.03 |
| $Li_3BS_3$ | -4.48 | -4.46 | -0.02 |
| $Li_5BSe_4$ | -3.72 | -3.76 | 0.04 |
| $Li_3BN_2$ | -5.41 | -5.56 | 0.15 |
| $Li_3AlF_6$ | -5.17 | -5.16 | -0.01 |
| $LiAlCl_4$ | -3.73 | -3.71 | -0.02 |
| $LiAlBr_4$ | -3.21 | -3.21 | 0 |
| $LiAlI_4$ | -2.75 | -2.75 | 0 |
| $Li_5AlO_4$ | -5.40 | -5.41 | 0.01 |
| $Li_5AlS_4$ | -4.14 | -4.16 | 0.02 |
| $Li_5AlSe_4$ | -3.78 | -3.79 | 0.01 |
| $Li_9AlN_4$ | -4.14 | -4.31 | 0.17 |
| $Li_2SiF_6$ | -5.18 | -5.24 | 0.06 |
| $LiSiCl_5$ | -3.63 | -3.64 | 0.01 |
| $LiSiBr_5$ | -3.10 | -3.13 | 0.03 |
| $LiSiI_5$ | -2.63 | -2.63 | 0 |
| $Li_4SiO_4$ | -5.87 | -5.88 | 0.01 |
| $Li_4SiS_4$ | -4.30 | -4.36 | 0.06 |
| $Li_4SiSe_4$ | -4.04 | -4.05 | 0.01 |
| $Li_8SiN_4$ | -4.12 | -4.23 | 0.11 |
| $LiPF_6$ | -4.86 | -4.90 | 0.04 |
| $LiPCl_6$ | -3.05 | -3.01 | -0.04 |
| $LiPBr_6$ | -2.63 | -2.62 | -0.01 |
| $LiPI_6$ | -1.85 | -1.83 | -0.02 |
| $Li_3PO_4$ | -6.15 | -6.14 | -0.01 |
| $Li_3PS_4$ | -4.23 | -4.27 | 0.04 |
| $Li_3PSe_4$ | -3.79 | -3.78 | -0.01 |
| $Li_7PN_4$ | -4.73 | -4.83 | 0.1 |

**Supplementary Table S2. The root mean square errors (RMSE) of MLIP models for amorphous structures.**

| Model | Force RMSE (eV/Å) | Energy RMSE (eV/atom) |
| --- | --- | --- |
| $LiBF_4$ | 0.04 | 0.0003 |
| $Li_4SiSe_4$ | 0.02 | 0.0007 |
| $LiSiCl_3Se$ | 0.03 | 0.0007 |

**Supplementary Table S3. Total number of atoms in supercells for AIMD simulations.**

| Structure | Total Number of Atoms in Supercell | Structure | Total Number of Atoms in Supercell |
|---|---|---|---|
| $LiBF_4$ | 72 | $Li_2SiF_6$ | 81 |
| $LiBCl_4$ | 60 | $LiSiCl_5$ | 63 |
| $LiBBr_4$ | 60 | $LiSiBr_5$ | 63 |
| $LiBI_4$ | 60 | $LiSiI_5$ | 63 |
| $Li_3BO_3$ | 77 | $Li_4SiO_4$ | 81 |
| $Li_3BS_3$ | 70 | $Li_4SiS_4$ | 81 |
| $Li_3BSe_3$ | 70 | $Li_4SiSe_4$ | 72 |
| $Li_6BN_3$ | 80 | $Li_8SiN_4$ | 91 |
| $Li_3AlF_6$ | 70 | $LiPF_6$ | 96 |
| $LiAlCl_4$ | 60 | $LiPCl_6$ | 80 |
| $LiAlBr_4$ | 60 | $LiPBr_6$ | 80 |
| $LiAlI_4$ | 60 | $LiPI_6$ | 64 |
| $Li_5AlO_4$ | 80 | $Li_3PO_4$ | 72 |
| $Li_5AlS_4$ | 80 | $Li_3PS_4$ | 72 |
| $Li_5AlSe_4$ | 80 | $Li_3PSe_4$ | 56 |
| $Li_9AlN_4$ | 84 | $Li_7PN_4$ | 84 |
| crystalline $LiAlCl_4$ | 96 | crystalline $LiAlCl_2Br_2$ | 96 |
| crystalline $LiAlBr_4$ | 96 | $LiSiCl_3Se$ | 54 |

**Supplementary Table S4. Total number of atoms in supercells for MLIP-MD simulations.**

| Structure | Total Number of Atoms in Supercell |
| --- | --- |
| $LiBF_4$ | 576 |
| $Li_4SiSe_4$ | 576 |
| $LiSiCl_3Se$ | 432 |

**Supplementary Note S1. The specific methodology for building and equilibrating amorphous Li-M-X structures.**

According to the free-volume model[1], amorphous materials typically have slightly lower density than their crystalline counterparts due to structural disorder. We set the initial density of each amorphous Li-M-X structure to 0.95 times that of its corresponding crystal counterpart or decomposition products. Also, by maintaining separation between M-X polyatomic groups, we try to create a controlled environment where observed ionic conductivity differences can be directly attributed to M-X pair characteristics rather than M-X-M chain connectivity effects.

To construct the initial structural models for $Li_{cx-m}M^{m+}[X^{x-}]_c$ ($m$ = 3, 4, 5 and $x$ = 1, 2, 3 for M element and X element respectively, $n$ is the saturated coordination number for M-X polyanionic groups in crystals, and the atomic ratio $c$ is defined as $n$ if $nx > m$ or as the largest integer not exceeding $m/x$+1 if $nx \leq m$ to ensure charge neutrality), we extract $MX_n$ polyanionic groups from crystalline structures in the Materials Project (MP) database[2] as structural unit A. An isolated Li atom is defined as structural unit B, while an isolated X anion is designated as structural unit C. Then, under the certain density, for every M-X pairing, we randomly distribute A units ($MX_n$ polyanionic groups) and B units (Li atoms) in a cubic box close to 10 Å$^2$ × 10 Å$^2$ × 10 Å$^2$ (For systems with significant oxidation-state disparities between M and X elements (i.e. $c > n$), c units (isolated $X^{x-}$ anions) also need to be added to achieve valence balance.) on the basis of ensuring that different atoms do not overlap with each other.

After that, we conducted structural equilibration steps in Figure 1b with different melting temperatures for different systems and make sure we obtain the fully balanced amorphous structures. First, we applied Processes 1 and 2 for all systems (Supplementary Figure S1), then compared both the free energies and structural characteristics (RDF profiles and coordination number curves) of the resulting materials (Supplementary Table S1 and Supplementary Figure S2). The results demonstrate that, apart from nitrides, amorphous materials attained structurally consistent configurations through both Process 1 and Process 2. During the melting stage of Process 2, even the oxides with higher melting points started melting and reached energy equilibrium at 1500 K (Supplementary Figure S3). This confirms that most structures (excluding nitrides) experienced melting and subsequent quenching in Process 2, thus producing rational amorphous configurations. This demonstrates that for systems other than nitrides, direct equilibration of initially disordered structures at 300 K is sufficient to generate amorphous structures indistinguishable from those produced by melting-and-quenching. For the nitrides exhibiting significant energy differences between Process 1 and Process 2, we performed melting at 1900 K (for $Li_3BN_2$ and $Li_9AlN_4$) and 2500 K (for $Li_8SiN_4$ and $Li_7PN_4$), respectively, followed by quenching. Successful melting and attainment of energy equilibrium at these elevated temperatures were confirmed by the data presented in Supplementary Figure S4. After the structural equilibration steps, the successful construction of physically reasonable amorphous configurations was confirmed when all structures exhibited continuous energy fluctuations without further minimization, indicating they had reached metastable states with lower energies and higher thermodynamic probabilities. After that, we conducted structural relaxation by density functional theory (DFT) to determine the equilibrium density for all amorphous systems. So far, we have successfully obtained the reasonable structures of 32 amorphous Li-M-X systems.

**Supplementary Note S2. The information about the training process for MLIP models and the MLIP-based MD simulation process.**

The MLIP models use the training sets of AIMD simulations of 60 ps at 800 K and 100 ps at 300 K for each unit-cells of $LiBF_4$, $Li_4SiSe_4$ and $LiSiCl_3Se$ amorphous structures, using parameters the same as the examples of by the NequIP software. Three distinct MLIP models were trained respectively. The evaluation results of MLIP models were put in Supplementary Table S2 to confirm their reliability. We conducted a 3000 ps MLIP-based MD simulation using the trained model at 300 K, and then took the last 2000 ps for MSD calculation in Supplementary Figure S9 and Supplementary Figure S13.

---

[1] Turnbull, D. & Cohen, M. H. Free-Volume Model of the Amorphous Phase: Glass Transition. *J. Chem. Phys.* **34**, 120-125 (1961).

[2] Ong, S. P.,Wang, L., Kang, B. & Ceder, G. Li−Fe−P−$O_2$ Phase diagram from first principles calculations. *Chem. Mater.* **20**, 1798 (2008).